\documentclass[showpacs,preprintnumbers,amsmath,amssymb,prb,aps,10pt]{revtex4-1}

\usepackage{graphicx}
\usepackage{subfigure}
\usepackage{dcolumn}
\usepackage{bm}

\begin{document}

\title{Spin nematic phase in (quasi-)one-dimensional frustrated magnet in strong magnetic field}

\author{A. V. Syromyatnikov}
 \email{syromyat@thd.pnpi.spb.ru}
\affiliation{Petersburg Nuclear Physics Institute NRC "Kurchatov Institute", Gatchina, St.\ Petersburg 188300, Russia}
\affiliation{Department of Physics, St.\ Petersburg State University, 198504 St.\ Petersburg, Russia}

\date{\today}

\begin{abstract}

We discuss spin-$\frac12$ one-dimensional (1D) and quasi-1D magnets with competing ferromagnetic nearest-neighbor $J_1$ and antiferromagnetic next-nearest-neighbor $J$ exchange interactions in a strong magnetic field $H$. It is well known that due to attraction between magnons quantum phase transitions (QPTs) take place at $H=H_s$ from the fully polarized phase to nematic ones if $J>|J_1|/4$. Such a transition at $J>0.368|J_1|$ is characterized by a softening of the two-magnon bound-state spectrum. Using a bond operator formalism we propose a bosonic representation of the spin Hamiltonian containing, aside from bosons describing one-magnon spin-1 excitations, a boson describing spin-2 excitations whose spectrum coincides at $H\ge H_s$ with the two-magnon bound-state spectrum obtained before. The presence of the bosonic mode in the theory that softens at $H=H_s$ makes the QPT consideration substantially standard. In the 1D case at $H<H_s$, we find an expression for the magnetization which describes well existing numerical data. Expressions for spin correlators are obtained which coincide with those derived before either in the limiting case of $J\gg|J_1|$ or using a phenomenological theory. In quasi-1D magnets, we find that the boundary in the $H$--$T$ plane between the fully polarized and the nematic phases is given by $H_s(0)-H_s(T)\propto T^{3/2}$. Simple expressions are obtained in the nematic phase for static spin correlators, spectra of magnons and the soft mode, magnetization and the nematic order parameter. All static two-spin correlation functions are short ranged with the correlation length proportional to  $1/\ln(1+|J_1|/J)$. Dynamical spin susceptibilities are discussed and it is shown that the soft mode can be observed experimentally in the longitudinal channel.

\end{abstract}


\maketitle

\section{Introduction}
\label{int}

Spin nematic states with multiple-spin ordering and without the conventional long-range magnetic order have attracted much attention in recent years. Two-spin nematic order can be generally described by the tensor \cite{andreev} $Q_{jl}^{\alpha\beta} = \langle S_j^\alpha S_l^\beta\rangle-\delta_{\alpha\beta} \langle {\bf S}_j {\bf S}_l \rangle /3$. The antisymmetric part of $Q_{jl}^{\alpha\beta}$ is related to the vector chirality $\langle {\bf S}_j \times{\bf S}_l \rangle$. The formation of the vector chiral spin liquid was anticipated a long time ago in 2D frustrated spin systems. \cite{sokol,*colman,*chub2} Such states have been obtained recently in a ring-exchange spin-$\frac12$ model at $T=0$, \cite{chir} and in classical frustrated spin systems at $T\ne0$. \cite{cinti,*sasha} The symmetric part of $Q_{jl}^{\alpha\beta}$ describes a quadrupolar order. In this instance one distinguishes the cases of $j=l$ and $j\ne l$. It has been known for a long time that the one-site ($j=l$) nematic state can be stabilized by the sufficiently strong biquadratic exchange $({\bf S}_1{\bf S}_2)^2$. \cite{bi} The interest in this mechanism of multiple-spin ordering stabilization has been revived recently in connection with experiments on cold atom gases \cite{zhou} and on the disordered spin system $\rm NiGa_2S_4$. \cite{ari,*mila} 

\begin{figure}
 \includegraphics[scale=0.9]{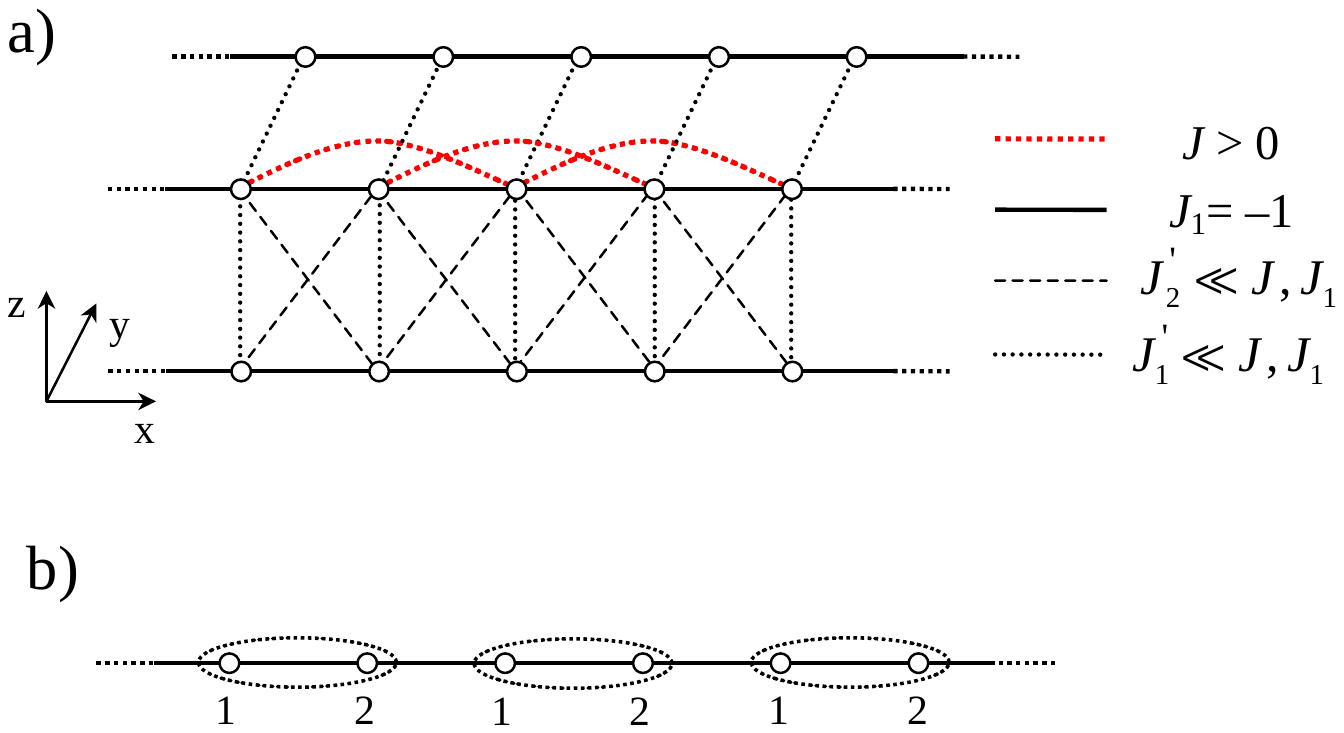}
 \caption{(Color online.) (a) Quasi-1D frustrated magnet described by Hamiltonian \eqref{ham}. It is implied that inter-chain interactions $J_1'$ and $J_2'$ are small compared to those inside chains. The approach suggested in the present paper is based on the unit cell doubling along chains direction that is shown in panel (b).}
 \label{chains}
\end{figure}

While the one-site quadrupolar ordering can exist only for $S\ge1$, the different-site one ($j\ne l$) can be found even in spin-$\frac12$ systems. In particular, the different-site nematic phases have been discussed recently in (quasi-)1D, \cite{chub,1d1,1d2,1d3,1d4,1d5,dmitr,sudan,kecke,kuzian,kuzian2,arlego,kolezh,zhito,ueda} 2D, \cite{2d0,2d1,2d2,2d3,referee1} and 3D \cite{3d} systems with competing ferro- and antiferromagnetic exchange couplings between neighboring and next-neighboring spins, respectively. The (quasi-)1D magnet of this kind is described by the Hamiltonian
\begin{equation}
\label{ham}
{\cal H} = \sum_j \left(-{\bf S}_j{\bf S}_{j+1} + J{\bf S}_j{\bf S}_{j+2}\right) - H\sum_j S^z_j + {\cal H}',
\end{equation}
where we set the ferromagnetic exchange coupling constant between neighboring spins to be equal to $-1$ and ${\cal H}'$ describes an inter-chain interaction that is also taken into account in the present paper (see Fig.~\ref{chains}(a)). As the field direction can be arbitrary, we direct the field perpendicular to chains for simplicity. The interest in model \eqref{ham} is stimulated also by recent experiments on the corresponding quasi-1D materials  LiCuVO$_4$, \cite{masuda,svistov,phaseli,nmrli,mour} $\rm Rb_2Cu_2Mo_3O_{12}$, \cite{rb} $\rm Li_2ZrCuO_4$, \cite{li} CuCl$_2$, \cite{cucl2} PbCuSO$_4$(OH)$_2$, \cite{pb} LiCuSbO$_4$, \cite{dutton} and some others.

It has been found recently that the physics of spin-$\frac12$ model \eqref{ham} with ${\cal H}'=0$ is even richer: field-driven transitions to phases with quasi-long-range multiple-spin ordering have been obtained below the saturation field $H_s$ and are described by operators $S_j^\pm S_{j+1}^\pm\dots S_{j+p-1}^\pm$ with $p=2$ (quadrupolar phase) at $J>0.368$, $p=3$ (hexapolar phase) at $0.284<J<0.368$ and $p=4$ (octupolar phase) at $0.259<J<0.284$. \cite{kecke,sudan} Finite inter-chain interaction ${\cal H}'$ stabilizes the long-range nematic order at $T=0$ but quite a small nonfrustrated ${\cal H}'$ turns the point $H=H_s$ into an ordinary quantum critical point separating the fully polarized phase and that with a long-range magnetic order. \cite{zhito,kuzian,kuzian2,ueda}

\begin{figure}
 \includegraphics[scale=0.95]{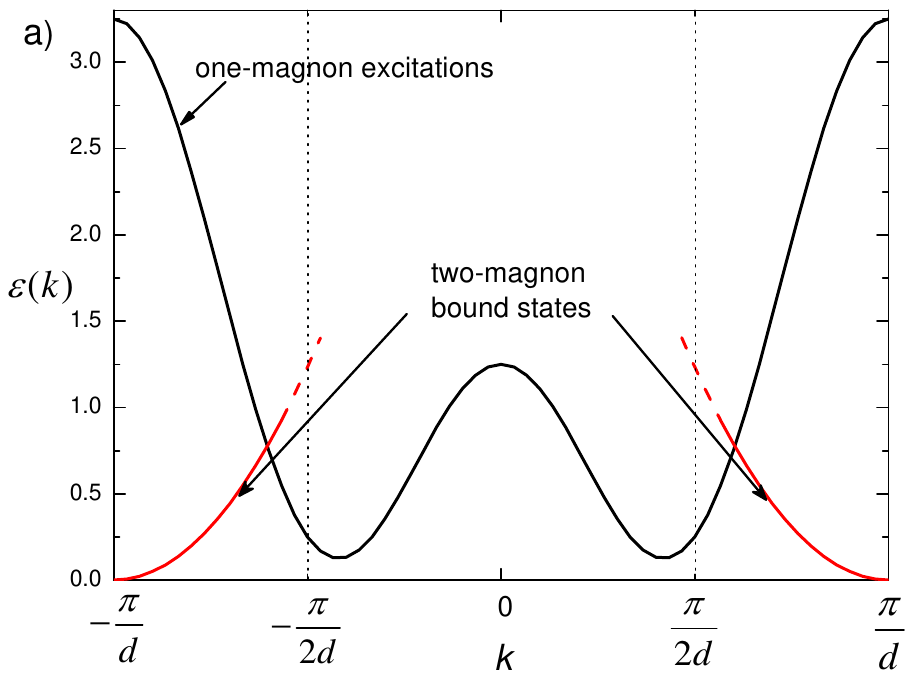}
 \includegraphics[scale=0.95]{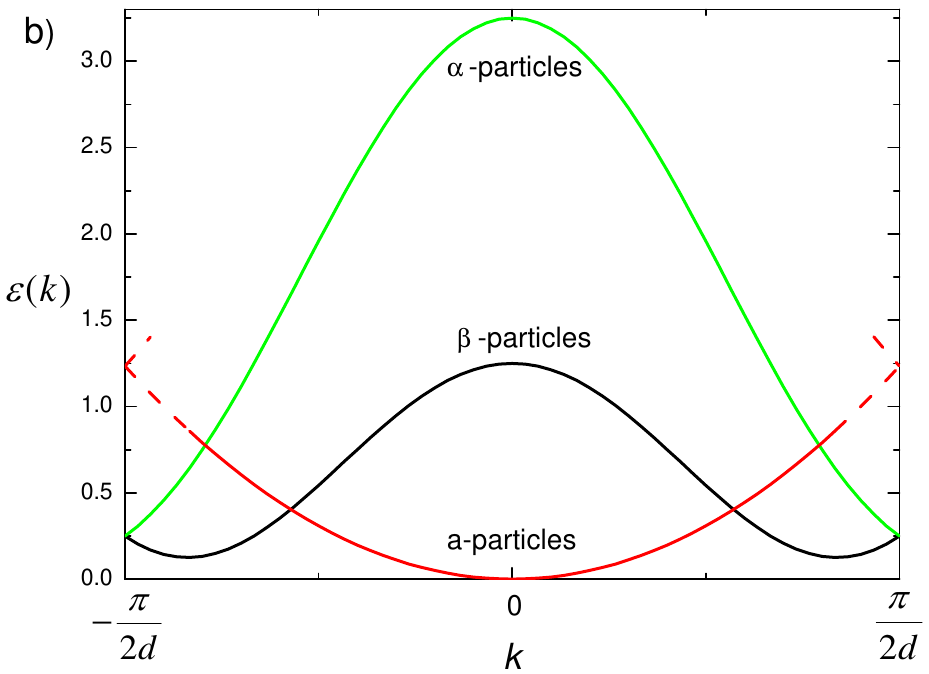}
 \caption{(Color online.) (a) Spectra of one-magnon excitations and two-magnon bound states are shown at $H=H_s$ as they were found before at $J>0.375$ for an isolated chain described by Hamiltonian \eqref{ham} (the particular curves drawn are for $J=1$). The distance between two neighboring spins in the chain is equal to $d$. Dashed lines are for parts of the bound-state spectrum which lie in the two-magnon continuum. The first Brillouin zone (BZ) $[-\pi/2d,\pi/2d]$ is marked after the unit cell doubling that is drawn in Fig.~\ref{chains}(b). The result of the reduction to this BZ is presented in panel (b). Within the approach suggested in the present paper, spectra of spin-1 excitations described by $\alpha$- and $\beta$- particles (see Eqs.~\eqref{h2}--\eqref{specb}) are two parts of the one-magnon spectrum shown in panel (a), whereas the spectrum of $a$-particles (spin-2 excitations) coincides with the low-energy part of the two-magnon bound-state spectrum.
 \label{appfig}}
\end{figure}

It is well known that the origin of the nematic phases is the attraction between magnons caused by frustration. \cite{chub} As a result of this attraction, the bottom of the one-magnon band lies above the lowest multi-magnon bound state at $H=H_s$ (see, e.g., Fig.~\ref{appfig}(a)). Then, transitions to nematic phases are characterized by a softening of the multi-magnon bound-state spectrum rather than the one-magnon spectrum. As a consequence, new approaches are required to describe such transitions. 

Properties of model \eqref{ham} are well-understood in the fully polarized state at $H\ge H_s$. Wave functions of the multi-magnon bound states can be represented as linear combinations of functions $S^-_{i_1}S^-_{i_2}\dots S^-_{i_p}|0\rangle$, where $|0\rangle$ is the vacuum state at which all spins have the maximum projection on the field direction, and the spectrum can be found numerically from the corresponding equations. \cite{mattis,kecke} An approach is suggested in Ref.~\cite{kuzian} that allows one to map the system at $H\ge H_s$ to a tight-binding impurity problem. In particular, it allows authors to obtain analytical expressions for the two-magnon bound-state spectrum and to show that it is quadratic at $J>0.375$ near its minimum located at $k=\pi/d$, where $d$ is the distance between neighboring spins (see Fig.~\ref{appfig}(a)). 

Transitions to nematic spin liquid phases in the purely 1D spin-$\frac12$ model are discussed using the bosonization technique in the limit $J\gg1$ and using a phenomenological approach at arbitrary $J>1/4$. \cite{1d1,1d2,1d3,1d5,kecke} According to the latter method the transition is equivalent to that in 1D hard-core Bose-gas with the following correspondence between the bosonic operators $b_i$ and spins: $b_i^\dagger = (-1)^iS^-_{i}S^-_{i+1}\dots S^-_{i+p-1}=(-1)^iM_i^{(p)}$ and $b_i^\dagger b_i = (1/2-S_i^z)/p$. Results of the phenomenological approach agree with those of the bosonization method in the region of its validity $J\gg1$ and show, in particular, an algebraic decay of static spin correlators $\langle S_i^zS_j^z\rangle$ and $\langle M_i^{(p)}M_j^{(p)}\rangle$ in nematic phases and an exponential decay of $\langle S_i^+S_j^-\rangle$. Although many predictions of the phenomenological theory are confirmed by numerical calculations, the corresponding microscopic analytical calculations based on the spin Hamiltonian are also desirable at $J\sim1$.

It is well known that the behavior of quasi-1D systems differs significantly at low $T$ from that of purely 1D systems. Then, a special approach for a quasi-1D model \eqref{ham} at $T=0$ has been suggested recently. \cite{zhito} The wave function of the ground state in the quadrupolar phase is proposed in a form that resembles the BCS pairing wave function of electrons in superconductors. Using this approach authors have calculated static spin correlators and found that in contrast to the purely 1D case $\langle S_i^zS_j^z\rangle$ decays exponentially and the system has a long-range nematic "antiferromagnetic" order. The magnetization of LiCuVO$_4$ at $H<H_s$ measured in Ref.~\cite{svistov} is also described successfully in Ref.~\cite{zhito}. However many dynamical properties as well as the temperature effect have not been considered yet in the quasi-1D case that leaves room for further theoretical discussion in this field.

We suggest an approach in the present paper that allows us to perform a quantitative microscopic consideration of the quadrupolar phase at $T\ge0$ and $H\approx H_s$ both in purely 1D and quasi-1D spin-$\frac12$ model \eqref{ham}. This approach is based on the unit cell doubling along the chain direction that is shown in Fig.~\ref{chains}(b) and on a representation of two spin operators in each unit cell via three bosons. This representation resembles those proposed for dimer spin-$\frac12$ systems. \cite{sachdev,chub89,kot} Two of these bosons describe one-magnon (spin-1) modes while the third one describes spin-2 excitations (see Fig.~\ref{appfig}) which are referred to as $a$-particles below. We demonstrate that the spectrum of $a$-particles coincides at $H\ge H_s$ with the spectrum of two-magnon bound states calculated before by other methods. \cite{chub,kuzian,kuzian2} Then, it is the main advantage of our approach that it contains a boson whose spectrum becomes "soft" as a result of the transition to the nematic phase. This circumstance makes relatively simple and quite standard the quantitative discussion of the transition.

It should be noted at once that the procedure of unit-cell doubling is arbitrary (there are two ways for neighboring spins to group into couples) and it breaks the initial translational symmetry. However, it does not play a role in our consideration because all the physical results are obtained in the present paper either {\it exactly} or using perturbation theories with "good" small parameters. As a consequence, the translational symmetry is restored in our results at $H\ge H_s$, it turns out to be broken at $H<H_s$ in accordance with conclusions of previous considerations and all the physical results obtained at $H<H_s$ do not depend on the way of spins grouping into couples.

By using our approach, we confirm the hypothesis proposed in Ref.~\cite{kecke} that the transition in an isolated chain to the quadrupolar phase is equivalent in many respects to that in 1D systems of hard-core bosons. We rederive many of the results for the isolated chain obtained in Refs.~\cite{1d1,1d2,1d3,1d5,kecke}. As an extension of the previous discussion, we derive an expression for the magnetization that describes well available numerical data at $H\approx H_s$.

In quasi-1D systems, we find that the boundary in the $H$--$T$ plane between the fully polarized and the quadrupolar phases is given by $H_s(0)-H_s(T)\propto T^{3/2}$. Simple expressions are obtained for static spin correlators, spectra of magnons and the soft mode, magnetization and the nematic order parameter. We obtain an "antiferromagnetic" nematic long-range order along the chains in accordance with Refs.~\cite{chub,zhito}. All the static two-spin correlators decay exponentially with the correlation length proportional to $1/\ln(1+1/J)$. This exponential decay results in broad peaks in the transverse structure factor with the period along chains equal to $\pi/d$ rather than $2\pi/d$. Dynamical spin susceptibilities $\chi_{\alpha\beta}(\omega,{\bf q})$ are discussed, where $\alpha,\beta=x,y,z$. It is shown that $\chi_{zz}(\omega,{\bf q})$ has sharp peaks at $\omega$ equal to energies of $a$-particles. Thus, the soft mode can be observed experimentally in the longitudinal channel. There are sharp peaks at $\omega$ corresponding to energies of magnons in transverse components of the dynamical spin susceptibility (in accordance with predictions of Ref.~\cite{zhito}). An application is discussed of the proposed theory to $\rm LiCuVO_4$. Our results are in reasonable agreement with available experimental data for magnetization in this compound.

The rest of the present paper is organized as follows. We describe in detail our approach in Sec.~\ref{approach}. Properties of an isolated chain are discussed in Sec.~\ref{1d}. Quasi-1D systems at $H\ge H_s$ and $H<H_s$ are considered in Secs.~\ref{quasi1d} and \ref{quasi1d<}, respectively. Sec.~\ref{conc} contains a summary of our results and a conclusion. Some details of calculations and the model describing LiCuVO$_4$ are discussed in appendices.

\section{Approach}
\label{approach}

\subsection{Spin representation}

We start with an isolated chain at $H\ge H_s$. To describe its properties in the fully polarized state and the quantum phase transition we suggest to double the unit cell as it is shown in Fig.~\ref{chains}(b). Then, there are two spins, ${\bf S}_{1j}$ and ${\bf S}_{2j}$, in $j$-th unit cell. To take into account all spin degrees of freedom in each unit cell we introduce three Bose-operators $a_j^\dagger$, $b_j^\dagger$ and $c_j^\dagger$ which create three spin states from the vacuum $|0\rangle$ as follows: 
\begin{eqnarray}
\label{states}
|0\rangle &=& \left|\uparrow\uparrow\rangle\right.,\nonumber\\
a_j^\dagger |0\rangle &=& \left|\downarrow\downarrow\rangle\right.,\nonumber\\
b_j^\dagger |0\rangle &=& \left|\uparrow\downarrow\rangle\right.,\\
c_j^\dagger |0\rangle &=& \left|\downarrow\uparrow\right.\rangle,\nonumber
\end{eqnarray}
where all spins have the maximum projection on the field direction at the state $|0\rangle$. One leads to the following spin representation via these Bose-operators:
\begin{align}
\label{trans}
S_{1j}^\dagger =& b_j^\dagger a_j + c_j, & S_{2j}^\dagger =& c_j^\dagger a_j + b_j,\nonumber\\
S_{1j}^- =& a_j^\dagger b_j + c_j^\dagger, & S_{2j}^- =& a_j^\dagger c_j + b_j^\dagger,\\
S_{1j}^z =& \frac12 -a_j^\dagger a_j - c_j^\dagger c_j, & S_{2j}^z =& \frac12 -a_j^\dagger a_j - b_j^\dagger b_j.\nonumber
\end{align}

It is easy to verify that Eqs.~\eqref{trans} reproduce spin commutation relations on the physical subspace (which consists of states with no more than one particle $a$, $b$ or $c$ in each unit cell) of the Hilbert space and ${\bf S}_{1j}^2={\bf S}_{2j}^2=3/4$. In order to eliminate contributions to physical quantities from unphysical states one can introduce into Eqs.~\eqref{trans} the projector operator $1-a_j^\dagger a_j-b_j^\dagger b_j-c_j^\dagger c_j$ or add to the Hamiltonian a term describing infinite repulsion between particles in each unit cell
\begin{equation}
\label{constr}
U\sum_j \left( a_j^\dagger a_j^\dagger a_j a_j + b_j^\dagger b_j^\dagger b_j b_j + c_j^\dagger c_j^\dagger c_j c_j + a_j^\dagger b_j^\dagger a_j b_j + a_j^\dagger c_j^\dagger a_j c_j + b_j^\dagger c_j^\dagger b_j c_j  \right),
\qquad U\to\infty.
\end{equation}
Both methods should lead to the same results at small $T$ (see, e.g., Ref.~\cite{siza}) and we choose the last one in the present paper.

Representation \eqref{states}--\eqref{constr} is an analog of the bond-operator representation suggested in Ref.~\cite{sachdev} and applied to systems with singlet ("dimerized") ground states. In principle, Eqs.~\eqref{states}--\eqref{constr} can be derived from Ref.~\cite{sachdev} implying that $\left|\uparrow\uparrow\right\rangle$ is the ground state as it is done, e.g., in Ref.~\cite{kot} for the singlet ground state.

\subsection{Hamiltonian transformation}

Substituting Eqs.~\eqref{trans} into Hamiltonian \eqref{ham} with ${\cal H}'=0$ and taking into account constraint \eqref{constr} one obtains
\begin{equation}
	{\cal H} = {\cal E}_0+ {\cal H}_2 +{\cal H}_3 +{\cal H}_4,
\end{equation}
where ${\cal E}_0$ is a constant,
\begin{eqnarray}
\label{h2}
{\cal H}_2 &=& \sum_{\bf k}\left(
\epsilon_{a0}a_{\bf k}^\dagger a_{\bf k} + 
E_{\bf k} \left(b_{\bf k}^\dagger b_{\bf k} + c_{\bf k}^\dagger c_{\bf k} \right)
+B_{\bf k} c_{\bf k}^\dagger b_{\bf k}
+B_{\bf k}^* b_{\bf k}^\dagger c_{\bf k}
\right),\\
\label{h3}
{\cal H}_3 &=& \frac{1}{\sqrt N}\sum_{{\bf k}_1+{\bf k}_2+{\bf k}_3={\bf 0}} 
\left(
-\frac14\left(e^{ik_1} + e^{ik_2}\right)c^\dagger_1c^\dagger_2a_{-3}
-\frac14\left(e^{-ik_1} + e^{-ik_2}\right)b^\dagger_1b^\dagger_2a_{-3}
+\frac12(J_1 + J_2)b^\dagger_1c^\dagger_2a_{-3}
\right.\nonumber\\
&&\left.
{}
-\frac14\left(e^{ik_2} + e^{ik_3}\right)a^\dagger_1c_{-2}c_{-3}
-\frac14\left(e^{-ik_2} + e^{-ik_3}\right)a^\dagger_1b_{-2}b_{-3}
+\frac12(J_2 + J_3)a^\dagger_1b_{-2}c_{-3}
\right),\\
\label{h4}
{\cal H}_4 &=& \frac1N\sum_{{\bf k}_1+{\bf k}_2+{\bf k}_3+{\bf k}_4={\bf 0}} 
\left(
\left(U+J_{1+3}-e^{i(k_1+k_3)}\right)a^\dagger_1a^\dagger_2a_{-3}a_{-4}
+\left( U+\frac12J_{1+3} \right) 
\left(b^\dagger_1b^\dagger_2b_{-3}b_{-4} + c^\dagger_1c^\dagger_2c_{-3}c_{-4}\right)
\right.\nonumber\\
&&{}
+\left(U+ J_{1+3} + \frac12J_{1+4}-e^{-i(k_1+k_3)} \right) a^\dagger_1c^\dagger_2a_{-3}c_{-4}
+\left(U+ J_{1+3} + \frac12J_{1+4}-e^{i(k_1+k_3)} \right) a^\dagger_1b^\dagger_2a_{-3}b_{-4}\nonumber\\
&&{}\left.
+\left(U - e^{-i(k_1+k_3)} \right) b^\dagger_1c^\dagger_2b_{-3}c_{-4}
- \frac12 e^{-i(k_2+k_3)} a^\dagger_1c^\dagger_2a_{-3}b_{-4}
- \frac12 e^{-i(k_1+k_4)} a^\dagger_1b^\dagger_2a_{-3}c_{-4}
\right),
\end{eqnarray}
${\bf k}$ is the one-dimensional momentum here, $J_{\bf k}=2J\cos k$, we set $2d=1$, $\epsilon_{a0}=2H+1-J_{\bf 0}$, $E_{\bf k}=H+1-(J_{\bf 0}-J_{\bf k})/2$, $B_{\bf k}=-e^{ik/2}\cos\frac k2$, $N$ is the number of unit cells (that is half the number of spins in the lattice), and we omit some indexes ${\bf k}$ in Eqs.~\eqref{h3} and \eqref{h4}. 

After the unitary transformation
\begin{equation}
\label{ut}
c_{\bf k}=\frac{1}{\sqrt2}e^{ik/4}\left(\alpha_{\bf k} + \beta_{\bf k}\right),
\quad
b_{\bf k}=\frac{1}{\sqrt2}e^{-ik/4}\left(\beta_{\bf k} - \alpha_{\bf k}\right),
\end{equation}
the bilinear part of the Hamiltonian \eqref{h2} acquires the form
\begin{equation}
\label{h22}
{\cal H}_2 = \sum_{\bf k} \left(
\epsilon_{a0}a_{\bf k}^\dagger a_{\bf k} 
+ \epsilon_\alpha({\bf k}) \alpha_{\bf k}^\dagger \alpha_{\bf k} + \epsilon_\beta({\bf k}) \beta_{\bf k}^\dagger \beta_{\bf k} 
\right),
\end{equation}
where
\begin{eqnarray}
\label{speca}
\epsilon_\alpha({\bf k}) &=& H+1-J+J\cos k + \cos\frac k2,\\
\label{specb}
\epsilon_\beta({\bf k}) &=& H+1-J+J\cos k - \cos\frac k2.
\end{eqnarray}
Eqs.~\eqref{speca} and \eqref{specb} represent two branches of the one-magnon spectrum which result from the unit-cell doubling as it is shown in Fig.~\ref{appfig}. The lower branch, $\epsilon_\beta({\bf k})$, has minima at ${\bf k}=(\pm Q,0,0)$, where 
\begin{equation}
\label{q}
\cos \frac Q2=\frac{1}{4J},
\end{equation}
near which $\epsilon_\beta({\bf k})$ is quadratic. As it is demonstrated below, the one-magnon branches are not renormalized at $H\ge H_s$ within our approach and Eqs.~\eqref{speca}--\eqref{specb} reproduce the one-magnon spectrum of the model \eqref{ham} obtained before \cite{chub,zhito} (remember that in our notation the distance between neighboring spins is equal to 1/2). In particular, one has from Eq.~\eqref{specb} for the field value at which the one-magnon spectrum becomes unstable
\begin{equation}
\label{hc}
H_c = 2J-1+\frac{1}{8J}.
\end{equation}

In contrast to $b$- and $c$- particles (or $\alpha$- and $\beta$- ones), $a$-particles are of the two-magnon nature (see Eqs.~\eqref{states}). It is one of the main findings of the present paper that their spectrum derived below coincides at $H\ge H_s$ with the low-energy part of the two-magnon bound-state spectrum obtained before by using other approaches \cite{chub,kuzian,kuzian2} (see Fig.~\ref{appfig}). 

\subsection{Diagram technique}

We find it more convenient not to use the unitary transformation \eqref{ut} in the following and to introduce four Green's functions 
\begin{eqnarray}
\label{gadef}
 G_a ( k ) &=& - i \langle a_k a^\dagger_k \rangle, \\
\label{gbdef}
 G_b ( k ) &=& - i \langle b_k b^\dagger_k \rangle, \\
\label{gcdef}
 G_c ( k ) &=& - i \langle c_k c^\dagger_k \rangle, \\
 \label{fdef}
 F ( k ) &=& - i \langle b_k c^\dagger_k \rangle, \\
\label{fndef}
 {\overline F} ( k ) &=& - i \langle c_k b^\dagger_k \rangle,
\end{eqnarray}
where $k=(\omega,{\bf k})$ and $a_k$ is the Fourier transform of $a_{\bf k}(t)$. Notice that $a$-particle can not transform to a single $b$- or $c$- particle due to the spin conservation law ($a$-particles carry spin 2 while $b$- and $c$- particles carry spin 1). That is why the Dyson equation for $G_a(k)$ has the form 
$
	G_a ( k ) = G_{a0} ( k ) ( 1 + \Sigma_a(k) G_a ( k ) ),
$
where $ G_{a0} ( k ) = ( \omega - \epsilon_{a0}+ i\delta )^{-1} $ and $\epsilon_{a0}$ is defined in Eq.~\eqref{h2} and which gives
\begin{equation}
	\label{ga}
	G_a ( k ) = \frac{1}{\omega-\epsilon_{a0}-\Sigma_a(k)+i\delta}.
\end{equation}
In contrast, $b$- and $c$- particles can interconvert that leads to two couples of Dyson equations for $G_b(k)$, $\overline{F}(k)$ and $G_c(k)$, $F (k)$. We obtain for one of them
\begin{equation}
\left\{
\label{dyson}
\begin{aligned}
 G_b ( k ) &= G_{b0} ( k ) \left( 1 + \Sigma_b(k) G_b ( k ) + \Pi(k) \overline{F} ( k ) \right), \\
 \overline{F} ( k ) &= G_{c0} (k) \left( \overline{\Pi}(k) G_b ( k ) + \Sigma_c(k) \overline{F} ( k ) \right),\\
\end{aligned}
\right.
\end{equation}
where $ G_{b0} ( k ) = G_{c0} ( k ) = ( \omega - E_{\bf k}+ i\delta )^{-1} $, and $\Sigma_{b,c}$, $\Pi$ and $\overline{\Pi}$ are self-energy parts. The solution of the Dyson equations has the form
\begin{eqnarray}
 \label{gb}
 G_b( k ) &=& \frac{\omega - E_{\bf k} - \Sigma_c(k)}{{\cal D}( k )}, \\ 
\label{gc}
 G_c( k ) &=& \frac{\omega - E_{\bf k} - \Sigma_b(k)}{{\cal D}( k )}, \\
 \label{f}
 \overline{F} ( k ) &=& \frac{ \overline{\Pi}(k)}{{\cal D}( k )}, 
\qquad
 F ( k ) = \frac{ \Pi(k)}{{\cal D}( k )}, \\
 \label{den} 
 {\cal D}( k ) &=& (\omega-E_{\bf k}-\Sigma_b(k)+i\delta)(\omega-E_{\bf k}-\Sigma_c(k)+i\delta) - \Pi(k)\overline{\Pi}(k).
\end{eqnarray}
It should be noted that all poles of these Green's functions lie below the real axis. This circumstance gives a useful rule of the diagram analysis at $H\ge H_s$: if a diagram contains a contour that can be walked around while moving by arrows of the Green's functions, integrals over frequencies in such a diagram give zero (see, e.g., Ref.~\cite{popov}). Using this rule one concludes that there are no nonzero diagrams for $\Sigma_b(k)$, $\Sigma_c(k)$ and $\Pi(k)$, $\overline{\Pi}(k)$ at $H\ge H_s$ which are determined solely by bilinear part of the Hamiltonian \eqref{h2}:
\begin{eqnarray}
\label{sigbc>}
\Sigma_b(k) &=& \Sigma_c(k)=0,\\
\label{pibc>}
\Pi(k) &=& \overline{\Pi}^*(k) =B_{\bf k}^*=-e^{-ik/2}\cos\frac k2.
\end{eqnarray}
As a result we have from Eqs.~\eqref{den}--\eqref{pibc>} 
${\cal D}( k ) = (\omega-\epsilon_\alpha({\bf k})+i\delta)(\omega-\epsilon_\beta({\bf k})+i\delta)$,
where $\epsilon_{\alpha,\beta}({\bf k})$ are given by Eqs.~\eqref{speca} and \eqref{specb}. Another useful rule for the diagram analysis follows from the spin conservation law. As the total spin carried by all incoming particles must be equal to that of all outgoing particles, one concludes immediately, for instance, that four-particle vertexes are zero with only one incoming or outgoing $a$-particle.

\begin{figure}
\includegraphics[scale = 0.73]{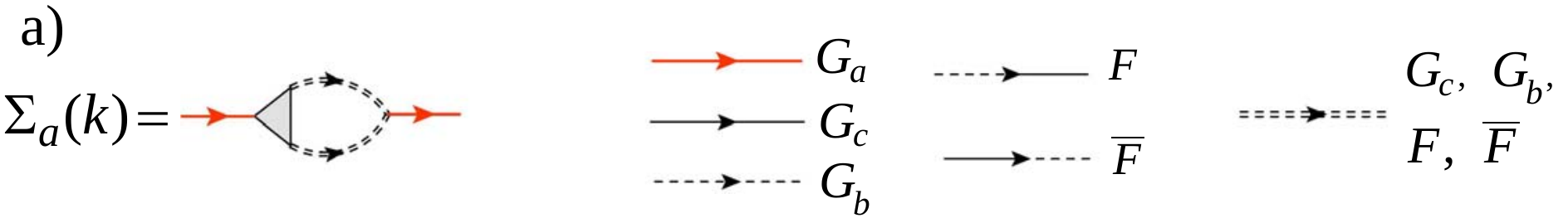}
\includegraphics[scale = 0.8]{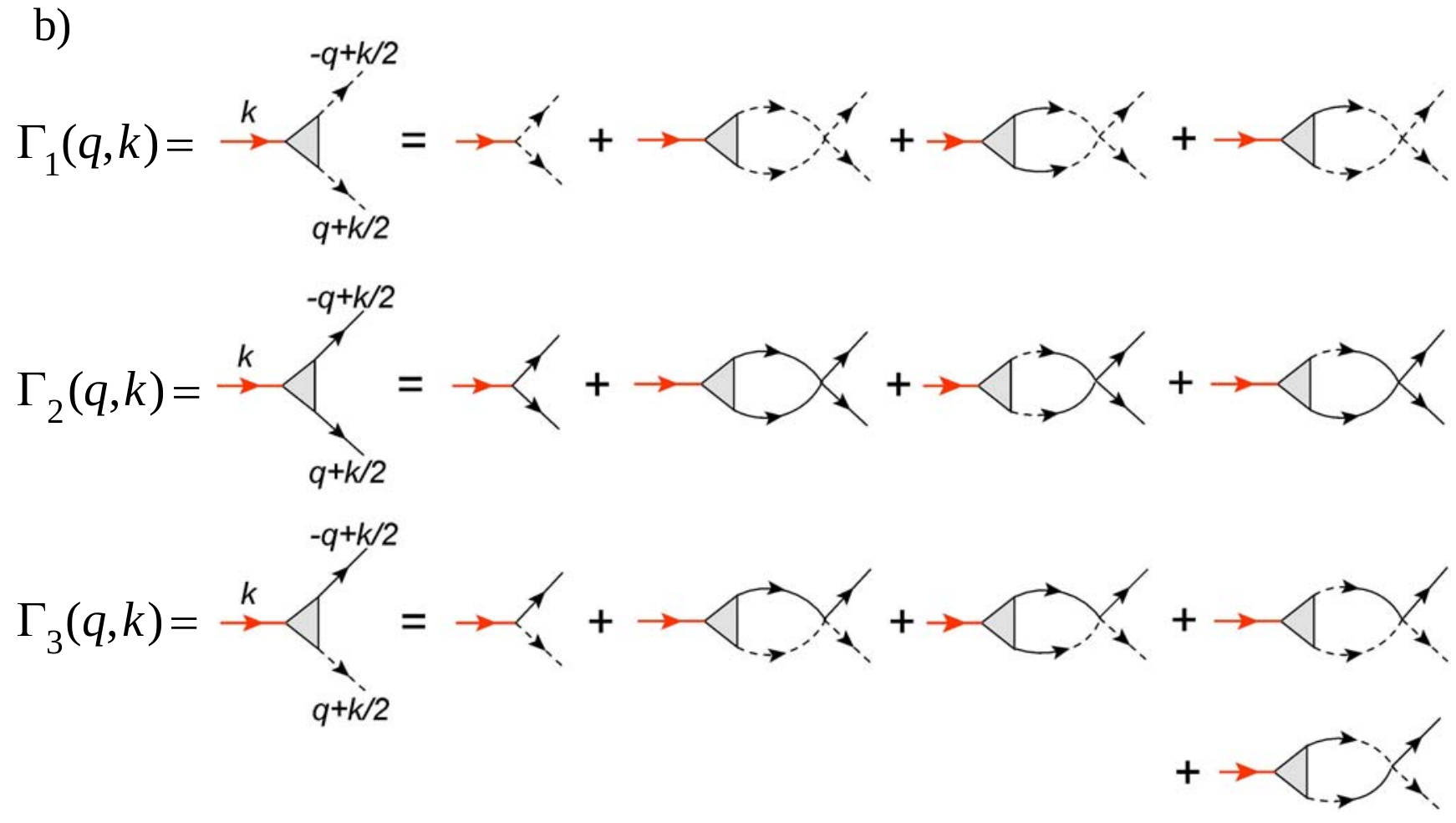}
\caption{(Color online.) (a) Diagrams for the self-energy part $\Sigma_a(k)$ of $a$-particles at $H\ge H_s$. It is implied that one should put Green's functions of $b$- and $c$- particles $G_b(p)$, $G_c(p)$, $F(p)$ and $\overline{F} ( p )$ defined by Eqs.~\eqref{gbdef}--\eqref{fndef} instead of double dashed lines (there are 10 different loop diagrams for $\Sigma_a(k)$). Triangles represent renormalized vertices for which we obtain equations presented in panel (b). Bare vertices are defined by Eqs.~\eqref{h3} and \eqref{h4}.}
\label{diaga}
\end{figure}

It is seen from Eq.~\eqref{h3} that the three-particle terms of the Hamiltonian describe decay of $a$-particles into two one-magnon particles of $b$ or $c$ types. They give rise to the only nonzero diagrams for $\Sigma_a(k)$ at $H\ge H_s$ all of which are shown in Fig.~\ref{diaga}.

\section{Isolated chain}
\label{1d}

\subsection{$H\ge H_s$}

To calculate $\Sigma_a(k)$ one has to derive three vertexes $\Gamma_{1,2,3}({\bf q},k)$ for which we obtain equations shown in Fig.~\ref{diaga}(b). The solution of these equations can be tried in the form
\begin{align}
\label{g12}
\Gamma_j({\bf q},k) &= u_j(k)+s_j(k)\cos q, & j=1,2,\\
\label{g3}
\Gamma_3({\bf q},k) &= u_3(k)+s_3(k)\cos q+iv(k)\sin q,&
\end{align}
where $u_{1,2,3}(k),s_{1,2,3}(k)$ and $v(k)$ are real functions. After substitution of Eqs.~\eqref{g12} and \eqref{g3} into equations shown in Fig.~\ref{diaga}(b), we obtain a set of seven linear algebraic equations for $u_{1,2,3}(k),s_{1,2,3}(k)$ and $v(k)$. The corresponding exact solution is quite cumbersome for arbitrary $k$. However, it simplifies greatly at $k=0$ and one has at $H=H_s$
\begin{eqnarray}
\label{g10}
\Gamma_1({\bf q},0) &=& \Gamma_2({\bf q},0) = - \cos^2\frac{q}{2},\\
\label{g30}
\Gamma_3({\bf q},0) &=& 2J-1+\frac{1}{1+J} + (1+2J)\cos q+i\sin q.
\end{eqnarray}
These expressions are used below for the nematic phase discussion. Substituting the general exact solution for the vertexes into expression for $\Sigma_a(k)$ shown in Fig.~\ref{diaga}(a) we lead to the following expression for the spectrum of $a$-particles at $k\ll1$:
\begin{eqnarray}
\label{ea}
\epsilon_a({\bf k}) &=& 2H+2-4J-\frac{1}{1+J} + D_\| k^2,\\
\label{dpar}
D_\| &=& \frac{3J^2(2+J)-1}{16(1+J)^2},
\end{eqnarray}
which coincides (up to a factor of 1/4 in $D_\|$ due to the unit cell doubling) with the spectrum of the two-magnon bound states obtained before within other approaches \cite{chub,kuzian}. Then, the condition $D_\|>0$ gives from Eq.~\eqref{dpar} that the minimum of the two-magnon bound-state spectrum is at $k=0$ if $J>0.375$ that is also in accordance with previous numerical findings \cite{chub,kecke} and analytical results \cite{kuzian}. It is not difficult within our approach to take into account also anisotropy in Hamiltonian \eqref{ham} of the form $S_j^zS_{j+1}^z$ and $S_j^zS_{j+2}^z$. We omit it in the present consideration for simplicity. The reader is referred to Ref.~\cite{kuzian} for the corresponding expressions for two-magnon bound-state spectra. 

The condition $\epsilon_a({\bf 0})=0$ gives from Eq.~\eqref{ea} the saturation field value for the isolated chain
\begin{equation}
\label{hs}
H_s = 2J-1+\frac{1}{2+2J}
\end{equation}
which is larger (if $J>1/3$) than the field value \eqref{hc} at which the one-magnon spectrum becomes gapless. We find also for the Green's function of $a$-particles near the pole
\begin{eqnarray}
\label{gapole}
G_a \left(\omega\approx\epsilon_a({\bf k}) ,{\bf k} \right) &=& \frac{Z}{\omega-\epsilon_a({\bf k})+i\delta},\\
\label{z}
Z &=& \frac{1+2J}{(1+J)(2J^2+2J+1)}.
\end{eqnarray}
These expressions are used in the subsequent calculations.

It should be noted that the bare completely flat spectrum of $a$-particles $\epsilon_{a0}$ is renormalized greatly by quantum fluctuations (see Eq.~\eqref{ea}). Because this renormalization comes from processes of $a$-particle decay into two one-magnon particles, one concludes that $a$-particle corresponds to a state that is a superposition of the initial state with two neighboring flipped spins and the great number of those with two flipped spins sitting on sites which can be quite far from each other. It is the structure of the wave function that is used in the standard method of the bound states analysis. \cite{mattis}

\subsection{$H<H_s$}

\begin{figure}
 \includegraphics[scale=0.82]{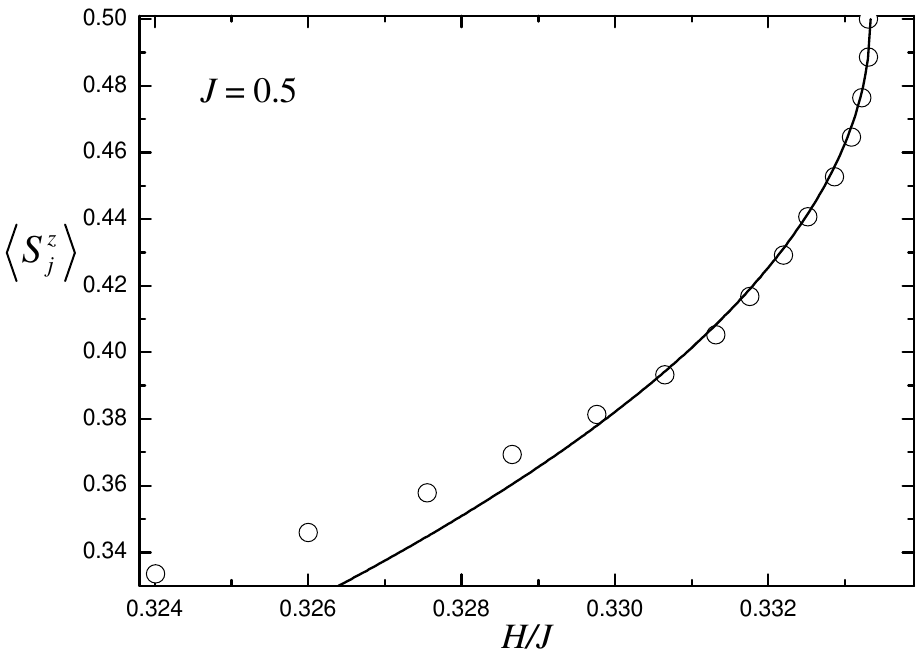}
 \caption{Magnetization of isolated chain with $J=0.5$. Circles are numerical data taken from Ref.~\cite{1d2} and the line is drawn using Eqs.~\eqref{mag1d}, \eqref{dpar} and \eqref{hs}.
 \label{mag1df}}
\end{figure}

The instability of $a$-particles spectrum \eqref{ea} at $H<H_s$ signifies the transition to the nematic phase. It is not the aim of the present paper to discuss in detail the quadrupolar phase in the isolated chain. But in view of the quadratic dispersion of $\epsilon_a({\bf k})$ at $H=H_s$ and gaps in spectra of $\alpha$- and $\beta$- particles, some results can be readily borrowed from the theory of 1D Bose gas of real particles. \cite{lieb,*lieb2,kor} In particular, one obtains $\langle a_j^\dagger a_j \rangle = \frac1\pi \sqrt{(H_s-H)/D_\|}$ using exact results by Lieb and Liniger \cite{lieb,*lieb2,kor} at $T=0$ and $H\approx H_s$. Diagram analysis shows that all poles of Green's functions $G_b(k)$ and $G_c(k)$ lie below the real axis at $H<H_s$ and only self-energy parts acquire some corrections (as is shown below, this is not the case in quasi-1D models due to the presence of the "condensate" of $a$-particles). Then, $\langle b_j^\dagger b_j \rangle=\langle c_j^\dagger c_j \rangle=0$ and one obtains from Eqs.~\eqref{trans} at $H\approx H_s$
\begin{equation}
\label{mag1d}
	\frac12 -  \left\langle S_j^z\right\rangle = \langle a_j^\dagger a_j \rangle = \frac1\pi \sqrt{\frac{H_s-H}{D_\|}}
\end{equation}
that is an extension of the previous isolated chain analysis. As it is demonstrated in Fig.~\ref{mag1df}, Eq.~\eqref{mag1d} describes well available numerical data \cite{1d2,1d3} within the range of its validity that reads as $\langle a_j^\dagger a_j \rangle\ll1$. Because $D_\|\ll H_s$ at $J>0.375$, the magnetization decays quite rapidly upon the field decreasing at $H\approx H_s$ that is also illustrated by Fig.~\ref{mag1df}.

Many results obtained before using other methods can be confirmed (and extended somewhat) within our approach. In particular, one obtains at $H\approx H_s$ from an expression for the asymptotic of the correlator of densities in 1D Bose-gas \cite{kor}
\begin{equation}
\label{zz}
\left\langle S_{j+n}^z(t) S_{j}^z(0) \right\rangle \approx \left\langle S_{j}^z \right\rangle^2 - \frac{1}{\pi}\left(\frac{1}{(n+iut)^2} + \frac{1}{(n-iut)^2}\right) 
+ A_1\frac{\cos(\pi\rho n)}{n^2+u^2t^2},
\end{equation}
where $n\to\infty$, $u=4\pi\rho D_\|$, $A_1$ is a constant and $\rho=\langle a_j^\dagger a_j \rangle$ is given by Eq.~\eqref{mag1d}. Eq.~\eqref{zz} coincides in the limit $H\to H_s$ with the corresponding expression derived in Refs.~\cite{1d1,1d2,kecke} using the bosonization technique ($J\gg1$) and the phenomenological theory of multipolar phases. Our discussion provides explicit expressions for the "sound velocity" $u$ and $\rho$ in Eq.~\eqref{zz}. Notice that spin-1 excitations do not contribute to Eq.~\eqref{zz} due to the above mentioned circumstance that all poles of one-magnon Green's functions lie below the real axis, whereupon all integrals over energies in the corresponding loop diagrams give zero at $T=0$. Then, the asymptotic of the 1D Bose gas "field correlator" gives for the nematic correlation at $H\approx H_s$ and $n\to\infty$
\begin{equation}
\label{nemcorr}
\left\langle S_{0}^+(t)S_{1}^+(t) S_{2n}^-(0) S_{2n+1}^-(0) \right\rangle 
=\langle a_0(t) a_n^\dagger(0) \rangle 
\approx \frac{A_2}{\sqrt{|2n+iut|}},
\end{equation}
where $A_2$ is a constant, that is consistent at $t=0$ with results of Refs.~\cite{1d5,1d2}.

It is seen from Eqs.~\eqref{zz} and \eqref{nemcorr} that static longitudinal and nematic correlators decay algebraically with the distance at $T=0$. According to the exact results of the 1D Bose gas theory \cite{kor} this algebraic decay changes into the following exponential decay at small $T$: $\left\langle S_{j+n}^z S_{j}^z \right\rangle \approx e^{-n/r_c}$ and $\left\langle S_{0}^+S_{1}^+ S_{2n}^- S_{2n+1}^- \right\rangle \approx e^{-n/2r_c}$, where $r_c\approx u/2\pi T$. Spin-1 excitations give contributions decaying much faster at small $T$ due to gaps in their spectra.
 
To conclude our brief discussion of the isolated chain at $H< H_s$, we confirm the long-standing expectation that bound states of two magnons behave in many respects like hard-core bosons. A more detailed consideration of the quadrupolar phase in the isolated chain is complicated by the existence of $b$- and $c$- particles. 

\section{Quasi-one-dimensional magnets. $H\ge H_s$.}
\label{quasi1d}

Let us take into account the inter-chain interaction. As it is shown below and as it was found before, \cite{zhito,kuzian,kuzian2,ueda} quite a small nonfrustrating interaction between chains can destroy the nematic phase and turn $H=H_s$ into an ordinary quantum critical point at which the condensation takes place of one-magnon excitations. That is why we consider in the present paper the inter-chain interaction as a perturbation. To demonstrate the main ideas we discuss the simplest interaction of the form (see Fig.~\ref{chains}(a))
\begin{equation}
\label{h'}
{\cal H}' = \frac12 \sum_{lm} J'_{1lm}{\bf S}_l{\bf S}_m  
= 
\frac12 \sum_{lm} \left(J'_{1lm}{\bf S}_{1l}{\bf S}_{1m} + J'_{1lm}{\bf S}_{2l}{\bf S}_{2m} \right),
\end{equation}
where the first sum is over the lattice sites and the second one is over the doubled unit cells shown in Fig.~\ref{chains}(b). Substituting Eqs.~\eqref{trans} into Eq.~\eqref{h'} one obtains
\begin{equation}
\label{h'k}
{\cal H}' = {\cal E}_0' +{\cal H}_2' +{\cal H}_3' +{\cal H}_4',
\end{equation}
where ${\cal E}_0'$ is a constant and
\begin{eqnarray}
\label{h2'}
{\cal H}_2' &=& \sum_{\bf k} \left(
-J_{1{\bf 0}}' a_{\bf k}^\dagger a_{\bf k} + 
\frac12\left(J_{1{\bf k}}'-J_{1{\bf 0}}'\right) 
\left( b_{\bf k}^\dagger b_{\bf k} + c_{\bf k}^\dagger c_{\bf k} \right)
\right),\\
\label{h3'}
{\cal H}_3' &=& \frac{1}{2\sqrt N}\sum_{{\bf k}_1+{\bf k}_2+{\bf k}_3=\bf 0} 
\left(
\left(J_{1{\bf k}_1}'+J_{1{\bf k}_2}'\right) b^\dagger_1c^\dagger_2a_{-3}
+\left(J_{1{\bf k}_2}'+J_{1{\bf k}_3}'\right) a^\dagger_1b_{-2}c_{-3}
\right),\\
\label{h4'}
{\cal H}_4' &=& \frac1N\sum_{{\bf k}_1+{\bf k}_2+{\bf k}_3+{\bf k}_4=\bf 0} 
\left(
J_{1{\bf k}_1+{\bf k}_3}' a^\dagger_1a^\dagger_2a_{-3}a_{-4}
+\frac12J_{1{\bf k}_1+{\bf k}_3}' 
\left( b^\dagger_1b^\dagger_2b_{-3}b_{-4} + c^\dagger_1c^\dagger_2c_{-3}c_{-4} \right)
\right.\nonumber\\
&&{}\left.
+\left( J_{1{\bf k}_1+{\bf k}_3}' +\frac12J_{1{\bf k}_1+{\bf k}_4}' \right) a^\dagger_1c^\dagger_2a_{-3}c_{-4}
+\left( J_{1{\bf k}_1+{\bf k}_3}' +\frac12J_{1{\bf k}_1+{\bf k}_4}' \right) a^\dagger_1b^\dagger_2a_{-3}b_{-4}
\right),
\end{eqnarray}
where all vectors $\bf k$ have two nonzero components, $y$ and $z$, (see Fig.~\ref{chains}(a)) and $J_{1 \bf k}'=\sum_{lm}J_{1lm}'e^{i({\bf k}{\bf R}_{lm})} = 2J_1'(\cos k_y+\cos k_z)$. Naturally, we will assume from now on that momenta ${\bf k}$ in Eqs.~\eqref{h2}--\eqref{h4} are 3D vectors with the only nonzero component $x$. It is easy to show that ${\cal H}'_3$ and ${\cal H}'_4$ do not lead to renormalization of the one-magnon spectrum at $H\ge H_s$. Then, one obtains for the spectrum of $\beta$-particles using Eqs.~\eqref{specb} and \eqref{h2'}
\begin{equation}
\label{alpha'}
\epsilon_\beta({\bf k}) = H+1-J+J\cos k_x - \cos\frac{k_x}{2} + \frac12\left(J_{1{\bf k}}'-J_{1{\bf 0}}'\right)
\end{equation}
that has minima at ${\bf k}=(\pm Q,0,0)$ and ${\bf k}=(\pm Q,\pi,\pi)$ if $J_1'<0$ and $J_1'>0$, respectively, and $Q$ is given by Eq.~\eqref{q}.

\begin{figure}
 \includegraphics[scale=0.82]{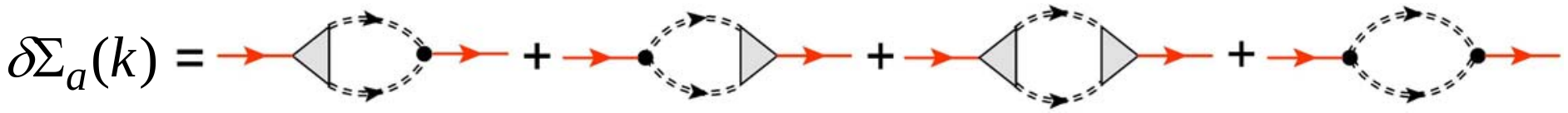}
 \caption{(Color online.) Diagrams for the self-energy part of $a$-particles which give contributions of the second order in the inter-chain interaction $J_1'$ at $H\ge H_s$. Triangles stand for vertexes which are calculated at $J_1'=0$ and equations for which are presented in Fig.~\ref{diaga}(b). Double dashed lines represent Green's functions $G_b$, $G_c$, $F$ and $\overline{F}$ which are calculated by taking into account also Eq.~\eqref{h2'}. Black dots stand for bare vertexes \eqref{h3'}.
 \label{disp}}
\end{figure}

In contrast, ${\cal H}'_3$ and ${\cal H}'_4$ renormalize the spectrum of $a$-particles $\epsilon_a({\bf k})$. Considering ${\cal H}'$ as a perturbation it is easy to demonstrate that the only diagrams contributing to $\epsilon_a({\bf k})$ in the second order in $J'_1$ are those presented in Fig.~\ref{disp}. Straightforward calculations show that their contribution to $\epsilon_a({\bf k})$ has the form 
\begin{eqnarray}
\label{dea'}
	\delta\epsilon_a({\bf k}) &=& -2D_\perp(2+\cos k_y+\cos k_z),\\
\label{dperp}
D_\perp &=& \left(J_1'\right)^2\frac{(1+J) (1+3 J+3 J^2)}{2(1+2 J)^2}.
\end{eqnarray}
Then, $\epsilon_a({\bf k})$ has a minimum at ${\bf k}=(0,0,0)$ near which we obtain using Eqs.~\eqref{ea} and \eqref{dea'}
\begin{eqnarray}
\label{ea'}
\epsilon_a({\bf k}) &=& 2H+2-4J-\frac{1}{1+J}-J_{1{\bf 0}}' -8D_\perp + D_\| k_x^2 +D_\perp k_\perp^2,
\end{eqnarray}
where $D_\|$ is given by Eq.~\eqref{dpar} and ${\bf k}_\perp$ is the projection of $\bf k$ on the $yz$ plane. It is seen from Eq.~\eqref{ea'} that $\epsilon_a({\bf k})$ is quadratic at $k\ll1$. Notice also that the stiffness $D_\perp$ of $a$-particles in the $yz$ plane is of the second order in $J_1'$.

One finds from Eqs.~\eqref{alpha'} and \eqref{ea'} for the field values at which spectra of $\beta$ and $a$-particles become gapless (cf. Eqs.~\eqref{hc} and \eqref{hs})
\begin{eqnarray}
\label{hc'}
H_c &=& 2J-1+\frac{1}{8J}+\frac12 \left|J_{1{\bf 0}}'\right| + \frac12 J_{1{\bf 0}}',\\
\label{hs'}
	H_s &=& 2J-1+\frac{1}{2+2J} + \frac12 J_{1{\bf 0}}'+4D_\perp.
\end{eqnarray}
The condition of the nematic phase existence ($H_s>H_c$) leads from Eqs.~\eqref{hc'} and \eqref{hs'} to the following inequality in the first order in $J_1'$:
\begin{equation}
\label{jcrit}
	\left|J_{1{\bf 0}}'\right| < \frac{3J-1}{4J(J+1)}.
\end{equation}
One concludes from Eq.~\eqref{jcrit} that $|J_1'|$ should be much smaller than unity in order the nematic phase can arise. Eqs.~\eqref{alpha'}, \eqref{hc'} and \eqref{hs'} are in accordance with previous results obtained within another approach. \cite{kuzian,kuzian2} The reader is referred to Ref.~\cite{kuzian2} for a more detailed discussion of the influence of the inter-chain interaction on properties of the considered system at $H\ge H_s$.

\section{Quasi-one-dimensional magnets. $H<H_s$.}
\label{quasi1d<}

It is convenient to discuss the quantum phase transition at $H=H_s$ in terms of Bose-condensation of $a$-particles that is similar to condensation of magnons in ordinary magnets in strong magnetic field. \cite{bat84}

\subsection{Condensation of $a$-particles}

\begin{figure}
 \includegraphics[scale=0.82]{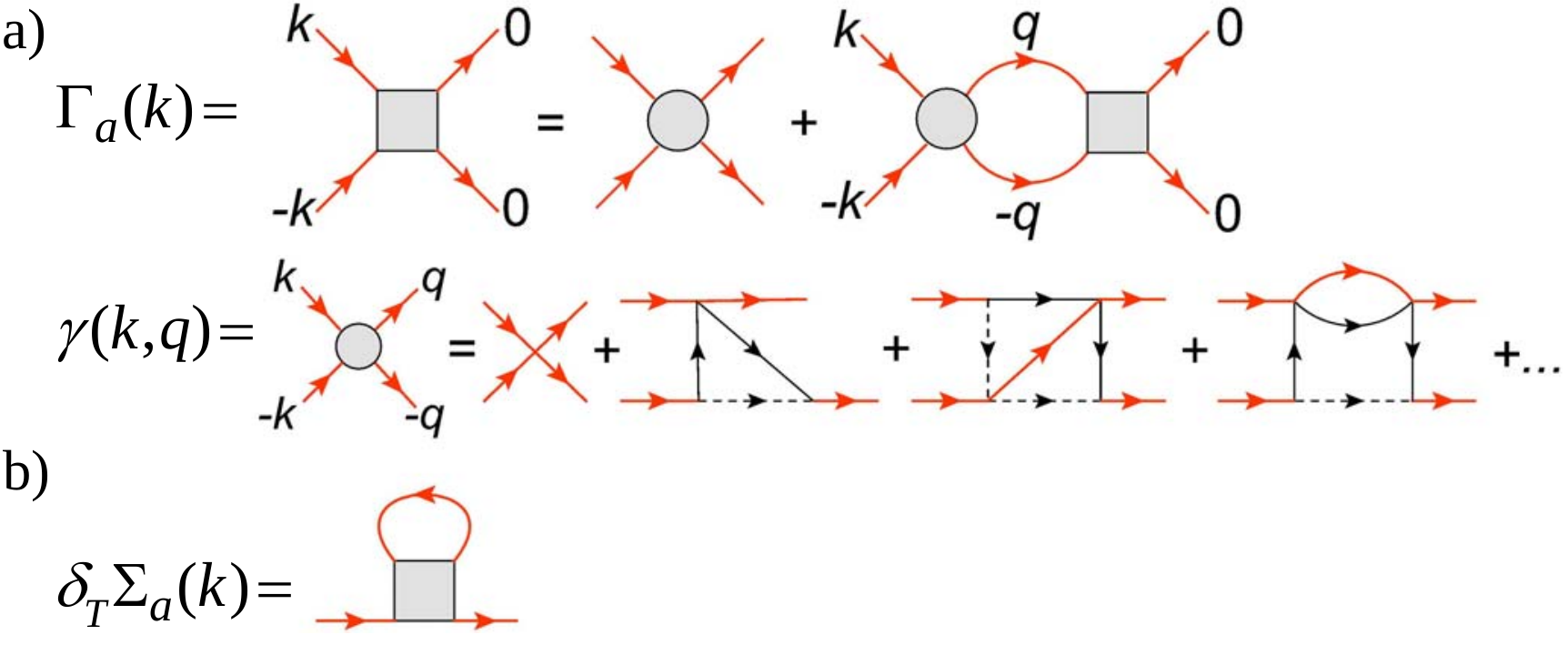}
 \caption{(Color online.) a) Equation for the four-particle vertex $\Gamma_a(k)$ of $a$-particles at $H\approx H_s$. A sum of an infinite number of diagrams which remain finite at $k,q\to0$ and $J_1'\to0$ plays the role of the "bare" vertex $\gamma(k,q)$. Some diagrams for $\gamma(k,q)$ are also presented. b) The diagram giving the leading temperature correction to the normal self-energy part of $a$-particles. Same notation as in Fig.~\ref{diaga}.
 \label{verta}}
\end{figure}

According to the general scheme \cite{popov,bat84}, one has to represent $a_{\bf 0}$ as follows at $H<H_s$:
\begin{equation}
\label{a0}
a_{\bf 0}\mapsto a_{\bf 0}+e^{i\phi}\sqrt{N\rho},
\end{equation}
where $\phi$ is an arbitrary phase and $\rho$ is the "condensate" density. New terms appear in the Hamiltonian after this transformation. In particular, the largest terms in the ground state energy have the form 
$
\delta {\cal E}_0 = -(H_s-H)\rho+\Gamma_a(0)\rho^2,
$
where $\Gamma_a(k)$ is the four-particle vertex of $a$-particles at $H=H_s$. An equation for $\Gamma_a(k)$ can be represented in the form shown in Fig.~\ref{verta}(a). Minimization of $\delta {\cal E}_0$ gives
\begin{equation}
\label{rho}
\rho = \frac{H_s-H}{2\Gamma_a(0)}.
\end{equation}
Although the above formulas are standard, \cite{popov} the situation is more complicated here. As it is shown in Fig.~\ref{verta}(a), one has to consider an infinite set of diagrams to find $\Gamma_a(k)$ which sum $\gamma(k,q)$ plays the role of a "bare" vertex. These diagrams remain finite at $k,q\to0$, $J_1'\to0$ because $\epsilon_{\alpha,\beta}({\bf k})$ have gaps at $H=H_s$ and $J_1'=0$. That is why one can set $J_1'=0$ in $\gamma(k,q)$. In contrast, the integral over $q$ diverges at $J_1'\to0$ in the second term on the right-hand side of the equation for $\Gamma_a(k)$ (see Fig.~\ref{verta}(a)). Due to the smallness of $J_1'$, small $q_x$ are important in the integral over $q$. As a result one obtains using Eq.~\eqref{gapole} the following equation in the leading order in $J_1'$: $\Gamma_a(0) = \gamma(0,0) - \gamma(0,0){\cal T}\Gamma_a(0)$ that gives
\begin{eqnarray}
\label{gammaa}
\Gamma_a(0) &=& \frac{\gamma(0,0)}{1+\gamma(0,0){\cal T}}\approx \frac{1}{\cal T},\\
\label{t}
	{\cal T} &=& \frac{Z^2}{(2\pi)^3}\int \frac{d{\bf k}}{\epsilon_a({\bf k})} \approx 0.32\frac{Z^2}{\sqrt{D_\|D_\perp}} \gg1,
\end{eqnarray}
where $D_\|$, $D_\perp$ and $Z$ are given by Eqs.~\eqref{dpar}, \eqref{dperp} and \eqref{z}, respectively, Eq.~\eqref{dea'} has been used to take the integral in Eq.~\eqref{t}, and we omit the unity in the denominator of Eq.~\eqref{gammaa} due to the large value of $\cal T$. We remind the reader also that the range of validity of Eqs.~\eqref{rho}--\eqref{t} is defined by the inequality
$
\rho\ll1.
$

According to the general theory of the dilute Bose gas condensation, \cite{popov,nep1,*nep2,pis,kop} the spectrum of $a$-particles acquires a linear part at small momenta $k$ and has the form
\begin{equation}
\label{eanem}
\widetilde{\epsilon}_a({\bf k}) = \sqrt{\epsilon_a({\bf k})(\epsilon_a({\bf k})+\Lambda)},
\end{equation}
where $\Lambda\sim\rho$ and $\epsilon_a({\bf k}) = D_\| k_x^2 +D_\perp k_\perp^2$.

Thermal fluctuations can be taken into account quite standardly. \cite{popov,nep1,*nep2,pis} The leading contribution from them to $\delta {\cal E}_0$ comes from the Hartree-Fock diagram shown in Fig.~\ref{verta}(b). As a result one obtains the following equation for the boundary in the $H$--$T$ plane between the fully polarized and the nematic phases at $T\ll D_\perp$:
\begin{equation}
\label{hst}
H_s(0) - H_s(T) = T^{3/2}\frac{Z\Gamma_a(0)}{D_\perp\sqrt{D_\|}} \frac{\zeta(3/2)}{2\pi^{3/2}},
\end{equation}
where $\Gamma_a(0)$ is given by Eq.~\eqref{gammaa}, $H_s(0)$ is given by Eq.~\eqref{hs'} and $\zeta(3/2)\approx2.61$ is the Riemann zeta-function. Notice that at small nonzero temperature one has to put $H_s(T)$ given by Eq.~\eqref{hst} instead of $H_s$ in Eq.~\eqref{rho}. 

\subsection{One-magnon spectrum renormalization}

Condensation of $a$-particles \eqref{a0} leads to terms in the Hamiltonian which renormalize the one-magnon spectrum. In particular, those coming from the three-particle vertexes and contributing to the bilinear part of the Hamiltonian \eqref{h2} have the form at $H\approx H_s$
\begin{eqnarray}
\label{dh2}
\delta {\cal H}_2^{(3)} = \sqrt\rho\sum_{\bf q}
&&\left(
e^{-i\phi}\Gamma_1({\bf q},0) \left(c_{\bf q}c_{-{\bf q}} + b_{\bf q}b_{-{\bf q}}\right)
+e^{i\phi}\Gamma_1({\bf q},0) \left(c^\dagger_{\bf q}c^\dagger_{-{\bf q}} + b^\dagger_{\bf q}b^\dagger_{-{\bf q}}\right)
\right.\nonumber\\
&&{}\left.
+e^{-i\phi}\Gamma_3({\bf q},0) b_{\bf q}c_{-{\bf q}}
+e^{i\phi}\Gamma_3^*({\bf q},0) b^\dagger_{\bf q}c^\dagger_{-{\bf q}}
\right),
\end{eqnarray}
where $\Gamma_1({\bf q},0)$ and $\Gamma_3({\bf q},0)$ are given by Eqs.~\eqref{g10} and \eqref{g30}, respectively, and we neglect the inter-chain interaction in the sum. There are also terms of the form $b^\dagger_{\bf q} b_{\bf q}$, $c^\dagger_{\bf q} c_{\bf q}$, $c^\dagger_{\bf q} b_{\bf q}$ and $b^\dagger_{\bf q} c_{\bf q}$ coming to ${\cal H}_2$ from four-particle vertexes $a^\dagger b^\dagger ab$, $a^\dagger b^\dagger ac$ and $a^\dagger c^\dagger ac$. The analysis of these vertexes similar to that carried out above for $\Gamma_a(k)$ shows that contributions to ${\cal H}_2$ from them are of the order of $\rho\sqrt{D_\perp}$. Straightforward calculations demonstrate that contribution from these terms to the spectrum renormalization is negligible. Thus, we use only Eq.~\eqref{dh2} below in order to make formulas more compact.

To find the one-magnon spectrum renormalization we take into account terms \eqref{dh2} in the Hamiltonian and introduce Green's functions 
\begin{equation}
\label{newgf}
P( k )= - i \langle b^\dagger_{-k} b^\dagger_k \rangle, \quad Q( k )= - i \langle c^\dagger_{-k} c^\dagger_k \rangle, \quad R( k )= - i \langle c^\dagger_{-k} b^\dagger_k \rangle
\end{equation}
in addition to those given by Eqs.~\eqref{gbdef}--\eqref{fndef}. Then, one leads to a set of four linear Dyson equations for $G_b$, $P$, $R$ and $\overline{F}$ that is derived and solved in Appendix~\ref{oneg}. Analysis of the Green's functions denominator shows that the spectrum of $\beta$-particles can be represented in the vicinity of its minimum as
\begin{align}
\label{teb}
\widetilde{\epsilon}_\beta({\bf k}) =& \sqrt{ \epsilon_\beta^2({\bf k}) -\rho A},\\
A =& \frac{(3 J-1) (J (2+J (19+8 J (J (16 J (3+4 J)-11)-9)))-1)}{128 J^4 (1+J)^2 (1+5 J)},
\end{align}
where $\epsilon_\beta({\bf k})$ is given by Eq.~\eqref{alpha'}, $A>0$ at $J>1/3$ and $A\approx12/5$ at $J\gg1$.

As it is seen from Eq.~\eqref{teb}, the spectrum of $\beta$-particles becomes unstable at a certain field value $\tilde H_c<H_s$ that would signify a transition to a phase with a long-range magnetic order. While the inequality $\rho\ll1$ can hold at $H=\tilde H_c$, the present discussion is not applicable for this quantum phase transition analysis because disappearance of the gap in the one-magnon spectrum leads to a finite damping of $a$-particles at {\it all} momenta. Besides, diagrams contributing to $\gamma(k,q)$ contains infrared divergences at $k,q\to0$ if the one-magnon spectrum is gapless. As a result the above analysis should be reconsidered at $H\approx\tilde H_c$, which is out of the scope of the present paper.

It should be noted also that one-magnon excitations acquire finite damping at $H<H_s$ stemming from loop diagrams. The damping, however, is small at $H\approx H_s$ being of the order of $\rho$ (that is much smaller than the gap in the spectra of $\alpha$- and $\beta$- particles). 

\subsection{Static spin correlators and nematic order parameter}

One obtains from Eqs.~\eqref{trans} and \eqref{a0} at $H<H_s$
\begin{eqnarray}
\left\langle {\bf S}_{1j}^\perp \right\rangle &=& \left\langle {\bf S}_{2j}^\perp \right\rangle =0,\\
\label{nemuc}
\left\langle S_{1j}^-S_{2j}^- \right\rangle &\equiv& \langle a_j^\dagger \rangle = \sqrt\rho e^{-i\phi},
\end{eqnarray}
where $\perp$ denotes the projection on the plane perpendicular to the field direction. Then, the condensation of $a$-particles signifies the formation of the quadrupolar phase without the conventional long-range magnetic order in which $\langle S_{1j}^-S_{2j}^- \rangle\ne0$ in each (double) unit cell. This condensate should appear also "between" the neighboring unit cells as the doubling of the unit cell we made is only a trick. To show this we calculate the value $\langle S_{2j}^-S_{1(j+1)}^- \rangle$ (see Fig.~\ref{chains}(b)), where $j$ enumerates sites in one of the chains, for which one has from Eqs.~\eqref{trans} in the leading order in $\rho$
\begin{equation}
\label{s21}
\left\langle S_{2j}^-S_{1(j+1)}^- \right\rangle = \frac1N\sum_{\bf q} e^{-iq_x} \left\langle b^\dagger_{\bf q} c^\dagger_{-{\bf q}} \right\rangle.
\end{equation}
One obtains after simple integration using Eqs.~\eqref{g10}, \eqref{g30}, \eqref{r} and \eqref{s21} in the leading order in the inter-chain interaction (cf.\ Eq.~\eqref{nemuc})
\begin{equation}
\label{s21f}
\left\langle S_{2j}^-S_{1(j+1)}^- \right\rangle = -\sqrt\rho e^{-i\phi}.
\end{equation}
As it is seen from Eqs.~\eqref{nemuc} and \eqref{s21f}, the nematic order parameter $\langle S_j^-S_{j+1}^- \rangle= (-1)^j\sqrt\rho e^{-i\phi}$, where $j$ enumerates now spins in a chain, has an "antiferromagnetic" order: its absolute value is the same for all $j$ whereas its phase differs by $\pi$ for two neighboring sites ($j$ and $j+1$). Such nematic ordering along chains was predicted in Refs.~\cite{chub,zhito}. In contrast, the nematic order is "ferromagnetic" in directions transverse to chains because the minimum of $\epsilon_a({\bf k})$ is at ${\bf k}=(0,0,0)$ (see Eqs.~\eqref{dea'} and \eqref{ea'}). The nematic ordering does not depend on the sign of $J_1'$. Notice also that this finding does not depend on the way of grouping of spins into couples shown in Fig.~\ref{chains}(b) due to arbitrariness of $\phi$.

\begin{figure}
 \includegraphics[scale=0.99]{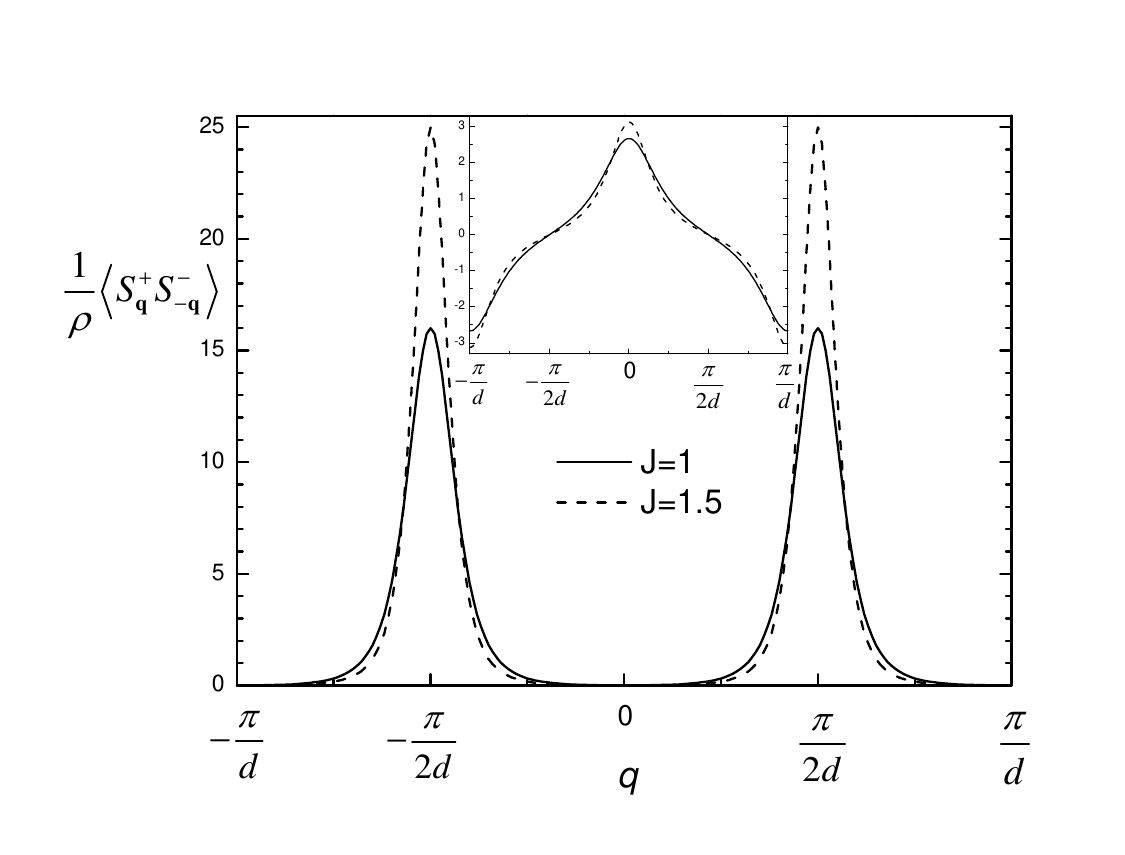}
 \caption{Static spin correlator at $H<H_s$ given by Eq.~\eqref{statxy} with ${\bf q} = (q,0,0)$. The inset shows the longitudinal static spin correlator divided by $\rho$ and given by Eq.~\eqref{statzz}. Here $d$ is the distance between two neighboring spins in a chain.
 \label{corr}}
\end{figure}

Let us consider the static spin correlator $\langle S_{j}^-S_{j+n}^- \rangle$. To find it one has to calculate $\langle S_{2j}^-S_{1(j+n)}^- \rangle$, $\langle S_{1j}^-S_{2(j+n)}^- \rangle$, $\langle S_{2j}^-S_{2(j+n)}^- \rangle$ and  $\langle S_{1j}^-S_{1(j+n)}^- \rangle$ that can be easily done in the first order in $\sqrt\rho$ and in the leading order in $J_1'$ using Eqs.~\eqref{trans}, \eqref{g10}, \eqref{g30}, \eqref{p} and \eqref{r}. The result can be represented in the form 
\begin{equation}
\label{nemorder}
\left\langle S_{j}^-S_{j+n}^- \right\rangle = \sqrt\rho e^{-i\phi} (-1)^j\sin\left(\frac{\pi n}{2}\right) \left(\frac{J}{1+J}\right)^{\frac{n-1}{2}},
\end{equation}
where $n>0$, that reproduces, in particular, Eqs.~\eqref{nemuc} and \eqref{s21f} at $n=1$. In much the same way, one obtains in the first order in $\rho$ and in the leading order in $J_1'$
\begin{align}
\label{spm}
\left\langle S_{j}^+S_{j+n}^- \right\rangle =& \rho \cos\left(\frac{\pi n}{2}\right) \left(\frac{J}{1+J}\right)^{\frac{n}{2}-1} \left( \frac{2J(1+J)}{1+2J} + \frac n2 \right),\\
\label{corrz}
\left\langle \left( S_{j}^z - \frac12 \right) \left( S_{j+n}^z- \frac12\right) \right\rangle =& 
\rho \sin^2\left(\frac{\pi n}{2}\right) \left(\frac{J}{1+J}\right)^{n-1}.
\end{align}
It is seen from Eqs.~\eqref{nemorder}--\eqref{corrz} that all the static spin correlators decay exponentially as $n\to\infty$. This should be contrasted with the case of the isolated chain in which the algebraic decay of static correlator \eqref{corrz} is observed. \cite{1d2,kecke} As a result of the exponential decay, static spin correlators have broad peaks in quasi-1D magnets instead of Bragg peaks which hight rises as $J$ increases. In particular, one obtains from Eqs.~\eqref{spm} and \eqref{corrz} up to a constant
\begin{align}
\label{statxy}
\left\langle S_{\bf q}^+S_{-{\bf q}}^- \right\rangle \sim & \rho\frac{\sin^2q}{4 J^2}\left(\frac{1}{4 J (1+J)}+\cos^2q\right)^{-2},\\
\label{statzz}
\left\langle S_{\bf q}^z S_{-{\bf q}}^z \right\rangle \sim& 
\rho \frac{1+2 J}{2 J^2} \cos q \left(\left(\frac{1}{2 J (1+J)}+1\right)^2-\cos^2q\right)^{-1},
\end{align}
where ${\bf q}=(q,0,0)$. Plots of these expressions are shown in Fig.~\ref{corr}. It is seen that the breakdown of the translational symmetry of the ground state at $H<H_s$ becomes apparent in the transverse spin structure factor which has a periodicity in the reciprocal space equal to $\pi/d$ rather than $2\pi/d$. 

A small inter-chain interaction gives rise to a weak dependence of correlators \eqref{statxy} and \eqref{statzz} on components of momenta transverse to chains. 
\footnote{
It should be pointed out also that Eq.~\eqref{statxy} is in agreement with the expression for the static transverse structure factor calculated in Ref.~\cite{zhito} for $J=2.375$ and $J_2'=-0.25$ (see Fig.~\ref{chains}(a)).
}

\subsection{Magnetization}

Using Eqs.~\eqref{trans} and the solution of Eqs.~\eqref{dyson2} for $G_b$ (and the similar expression for $G_c$) one finds after simple integration for the magnetization in the leading order in $\rho$ and $J_1'$ 
\begin{equation}
\label{mag}
	\langle S^z_j \rangle = \frac12 - 2\frac{(1+J)^2 }{1+2 J} \rho.
\end{equation}
Notice that in contrast to 1D case discussed above, there is a nonzero contribution to the second term in Eq.~\eqref{mag} from $\langle c_j^\dagger c_j \rangle=\langle b_j^\dagger b_j \rangle$.

\subsection{Spin Green's functions}

Spin Green's functions (or generalized susceptibilities) defined as
\begin{equation}
\label{sgf}
\chi_{\alpha\beta}(\omega,{\bf q}) = \left\langle S_{-\bf q}^\alpha, S_{\bf q}^\beta\right\rangle_\omega
=
i\int_0^\infty dt e^{i\omega t}\left\langle\left[S_{-\bf q}^\alpha(t), S_{\bf q}^\beta(0)\right]\right\rangle,
\end{equation}
where $\alpha,\beta={x,y,z}$, can be found straightforwardly using Eqs.~\eqref{trans} and \eqref{gb}--\eqref{pibc>}. In particular, components of $\chi_{\alpha\beta}$ transverse to the field direction are expressed via Green's functions $G_b$, $G_c$, $F$ and $\overline{F}$. As a consequence, they have sharp peaks at $\omega=\pm\tilde\epsilon_{\alpha,\beta}({\bf q})$ corresponding to one-magnon excitations. It is seen from Eqs.~\eqref{trans} that the longitudinal spin Green's function has a contribution containing $G_a(q)$ (apart from a smooth background originating from terms $c^\dagger c$, $b^\dagger b$ and $a^\dagger a$ in $S_1^z$ and $S_2^z$):
\begin{equation}
\label{chizz}
\chi_{zz}(\omega,{\bf q}) \sim - \rho (G_a(\omega,{\bf q}) + G_a(-\omega,{\bf q})) \approx -\frac{2Z\rho\tilde\epsilon_a({\bf q})}{(\omega+i\delta)^2-\tilde\epsilon^2_a({\bf q})}
\end{equation}
that shows sharp peaks at $\omega=\pm\tilde\epsilon_a({\bf q})$ and small $q$. Thus, the soft mode \eqref{eanem} can be observed in the nematic phase experimentally in the longitudinal channel.

\subsection{Phase transitions at $H<H_s(T)$. Symmetry consideration.}

Let us discuss the breakdown of symmetry in the transition from the fully polarized phase to the nematic one at $H=H_s(T)$. The symmetry of the Hamiltonian \eqref{ham} is $O(2)$. Symmetry operations which do not change the nematic order parameter $\langle S_j^-S_{j+1}^- \rangle= (-1)^j\sqrt\rho e^{-i\phi}$ include a rotation by $\pi$ as well as rotations by $-\phi$ and $-\phi+\pi$ accompanied by a reflection. These operations form a discrete group which is equivalent to $Z_2\otimes Z_2$. Then, the phase transition in the quadrupolar phase corresponds to the continuous symmetry $O(2)/(Z_2\otimes Z_2)=SO(2)/Z_2$ breakdown (see, e.g., Ref.~\cite{mermin}) that produces, in particular, the massless excitations (the soft mode). On the other hand one can expect in the considered quasi-1D system that the broken symmetry at small field is $Z_2\otimes SO(2)$ (as in a non-collinear Heisenberg $XY$-magnet). Consequently, the discrete subgroup $Z_2\otimes Z_2$ should be broken upon the field decreasing at $H<H_s(T)$. This breakdown of symmetry can happen in one transition (which can be either of the first or of the second order) or in two subsequent Ising (second order) transitions corresponding to two $Z_2$ subgroup breaking.
\footnote{We do not consider here the possibility of the symmetry restoration in transitions at $H<H_s(T)$ that can increase the number of phase transitions.} 
The latter scenario can be realized in LiCuVO$_4$, where two phase transitions (apart from the low-field spin-flop one) are observed at $H<H_s(T)$. \cite{masuda,svistov,phaseli,nmrli,mour}

\section{Summary and conclusion}
\label{conc}

To summarize, we suggest an approach for quantitative discussion of quantum phase transitions to the quadrupolar phase in frustrated spin systems in strong magnetic field $H\approx H_s$. Quasi-1D and 1D spin-$\frac12$ models described by Hamiltonian \eqref{ham} are discussed in detail. The approach we propose is based on the unit cell doubling along the chain direction presented in Fig.~\ref{chains}(b) and on representation \eqref{trans} of spins in each unit cell via three Bose-operators $a$, $b$ and $c$ \eqref{states}. Bosons $b$ and $c$ describe spin-1 excitations which spectra represent at $H\ge H_s$ two parts of the one-magnon spectrum (see Fig.~\ref{appfig}, Eqs.~\eqref{h2} and \eqref{ut}--\eqref{specb}). Spectra \eqref{speca} and \eqref{dea'}--\eqref{ea'} of the boson $a$ carrying spin 2 coincide at $H\ge H_s$ with those of two-magnon bound states found before within other approaches. It is the main advantage of the suggested approach that there is the bosonic mode in the theory softening at $H=H_s$. This circumstance makes the consideration of the transition to the quadrupolar phase substantially standard. In the purely 1D case, we rederive spin correlators \eqref{zz} and \eqref{nemcorr} obtained before either in the limiting case of $J\gg1$ or using the phenomenological theory and extend previous discussions by Eq.~\eqref{mag1d} for magnetization that describes well existing numerical data at $H\approx H_s$ (see Fig.~\ref{mag1df}). In the quasi-1D model with the simplest inter-chain interaction \eqref{h'}, we calculate at $H<H_s$ spectra of the one-magnon band \eqref{teb} and the soft mode \eqref{eanem}, nematic order parameter \eqref{nemuc} and \eqref{s21f}, static spin correlators \eqref{nemorder}--\eqref{statzz} and magnetization \eqref{mag} which are expressed via the condensate density $\rho$ given by Eqs.~\eqref{rho}--\eqref{t}. At $T\ne0$, $\rho$ is also expressed by Eqs.~\eqref{rho}--\eqref{t} in which one should replace $H_s$ by $H_s(T)$ given by Eq.~\eqref{hst}. All static two-spin correlators decay exponentially with the correlation length proportional to $1/\ln(1+1/J)$. This decay results in broad peaks in static structure factors (see Fig.~\ref{corr}). The periodicity in the reciprocal space of the transverse static structure factor is equal to $\pi/d$ rather than $2\pi/d$. Transverse components of the dynamical spin susceptibilities \eqref{sgf} are expressed via one-magnon Green's functions and contain sharp peaks at $\omega$ equal to magnon energies. The longitudinal component $\chi_{zz}(\omega,{\bf q})$ apart from a smooth background contains a contribution \eqref{chizz} from Green's functions of $a$-particles which have sharp peaks at $\omega=\pm\tilde\epsilon_a({\bf q})$. 

We apply the proposed approach to the analysis of the model describing the quasi-1D material LiCuVO$_4$ in which a transition at the saturation field to (presumably) a quadrupolar phase has been observed experimentally. Details of the corresponding calculations are presented in Appendix~\ref{licuvo}. Predictions of our theory are in reasonable agreement with the recent magnetization measurements in LiCuVO$_4$. Our finding that $H_s(0)-H_s(T)\propto T^{3/2}$, Eqs.~\eqref{eanem}, \eqref{teb}, \eqref{nemorder}--\eqref{mag} and \eqref{chizz} can be checked in further experiments on this material which should confirm also that the phase observed just below $H_s(T)$ is really the quadrupolar one.

\begin{acknowledgments}

This work was supported by the President of the Russian Federation (grant MD-274.2012.2), the Dynasty foundation and RFBR grants 12-02-01234 and 12-02-00498.

\end{acknowledgments}

\appendix

\section{One-magnon Green's functions in the nematic phase}
\label{oneg}

Taking into account terms \eqref{dh2} in the Hamiltonian which appear at $H<H_s$, one leads to a set of four linear Dyson equations for $G_b(k)$, $P(k)$, $R(k)$ and $\overline{F}(k)$ that reads (cf.~Eqs.~\eqref{dyson})
\begin{equation}
\left\{
\label{dyson2}
\begin{aligned}
 G_b ( k ) &= G_{b0} ( k ) \left( 1 + e^{i\phi}2\sqrt\rho\Gamma_1({\bf k},0) P ( k ) + e^{i\phi}\sqrt\rho\Gamma_3^*({\bf k},0) R ( k ) + B^*_{\bf k} \overline{F} ( k ) \right), \\
 P( k ) &= {\overline G}_{b0} ( k ) \left( e^{-i\phi}2\sqrt\rho\Gamma_1({\bf k},0) G_b( k ) + e^{-i\phi}\sqrt\rho\Gamma_3^*({\bf k},0) \overline{F} ( k ) + B^*_{\bf k} R( k ) \right), \\
R( k ) &= {\overline G}_{c0} ( k ) \left( e^{-i\phi}\sqrt\rho\Gamma_3({\bf k},0) G_b( k ) + B_{\bf k} P( k ) + e^{-i\phi}2\sqrt\rho\Gamma_1({\bf k},0) \overline{F} ( k ) \right), \\
 \overline{F} ( k ) &= G_{c0} (k) \left( B_{\bf k} G_b ( k ) + e^{i\phi}2\sqrt\rho \Gamma_1({\bf k},0) R( k ) + e^{i\phi}\sqrt\rho \Gamma_3({\bf k},0) P( k ) \right),\\
\end{aligned}
\right.
\end{equation}
where $P(k)$ and $R(k)$ are defined in Eqs.~\eqref{newgf}, $ G_{b0} ( k ) = G_{c0} ( k ) = \overline{G}_{b0} ( -k ) = \overline{G}_{c0} ( -k ) = ( \omega - E_{\bf k}'+ i\delta )^{-1} $, $E_{\bf k}'=E_{\bf k}+(J_{1{\bf k}}'-J_{1{\bf 0}}')/2$, $E_{\bf k}$ and $B_{\bf k}$ are defined in Eq.~\eqref{h2}. Solving Eqs.~\eqref{dyson2} one finds, in particular, in the leading order in $\rho$
\begin{eqnarray}
\label{p}
P( k ) &=& \sqrt{\rho}e^{-i\phi}\frac{2\Gamma_1({\bf k},0)  \left(E_{\bf k}^{\prime2}+|B_{\bf k}|^2-\omega ^2\right)-\Gamma_3^*({\bf k},0)B_{\bf k} (E_{\bf k}' +\omega )- B_{\bf k}^* \Gamma_3({\bf k},0)(E_{\bf k}' -\omega )}{(\omega-\epsilon_\alpha({\bf k})+i\delta)(\omega+\epsilon_\alpha({\bf k})-i\delta)(\omega-\epsilon_\beta({\bf k})+i\delta)(\omega+\epsilon_\beta({\bf k})-i\delta)},\\
\label{r}
R( k ) &=& \sqrt{\rho}e^{-i\phi}\frac{\Gamma_3^*({\bf k},0)B_{\bf k}^2-(\omega^2-E_{\bf k}^{\prime2})\Gamma_3({\bf k},0)-4E_{\bf k}'B_{\bf k}\Gamma_1({\bf k},0)}{(\omega-\epsilon_\alpha({\bf k})+i\delta)(\omega+\epsilon_\alpha({\bf k})-i\delta)(\omega-\epsilon_\beta({\bf k})+i\delta)(\omega+\epsilon_\beta({\bf k})-i\delta)},
\end{eqnarray}
where we set $\rho=0$ in the denominators. Expressions for $G_b$ and $\overline{F}$ are cumbersome and we do not present them here. 

Another set of Dyson equations for $G_c(k)$, $Q(k)$, $R(k)$ and $F(k)$ can be considered in much the same way.

\section{Application to LiCuVO$_4$}
\label{licuvo}

We apply in this appendix the approach proposed in the main text to the particular quasi-1D compound LiCuVO$_4$. While exchange coupling constants inside a chain $J_1=-18.5$~K and $J=44$~K have been extracted from experimental data in LiCuVO$_4$ quite precisely, small inter-chain interaction has been determined much less accurately. \cite{ender,svistov} Then, we take into account to first approximation only the inter-chain coupling $J_2'=-4.3$~K that is shown in Fig.~\ref{chains}(a). \cite{ender,svistov} Such a model for LiCuVO$_4$ is considered in Ref.~\cite{zhito} using another approach that involves numerical calculations at $H\ge H_s$ and self-consistent calculations at $H<H_s$. Thus, it is reasonable to compare our analytical results with those of Ref.~\cite{zhito}. Notice that the inter-chain interaction $J_2'$ makes the system two-dimensional (in contrast to $J_1'$ considered in the main text).

Let us start with the case of $H\ge H_s$ and represent the inter-chain interaction in the form
\begin{equation}
\label{h'2}
{\cal H}' = \frac12 \sum_{lm} J'_{2lm}{\bf S}_l{\bf S}_m 
= 
\sum_{lm} J'_{2lm}{\bf S}_{1l}{\bf S}_{2m},
\end{equation}
where the first sum is over the lattice sites and the second one is over the double unit cells shown in Fig.~\ref{chains}(b). Substituting Eqs.~\eqref{trans} into Eq.~\eqref{h'2} one obtains Eq.~\eqref{h'k}, where now
\begin{eqnarray}
\label{h2'2}
{\cal H}_2' &=& \sum_{\bf k} \left(
-J_{2{\bf 0}}'a_{\bf k}^\dagger a_{\bf k} -
J_{2{\bf 0}}' \frac12 
\left( b_{\bf k}^\dagger b_{\bf k} + c_{\bf k}^\dagger c_{\bf k} \right)
+\frac12J_{2{\bf k}}' c_{\bf k}^\dagger b_{\bf k}
+\frac12J_{2{-\bf k}}' b_{\bf k}^\dagger c_{\bf k}
\right),\\
\label{h3'2}
{\cal H}_3' &=& \frac{1}{2\sqrt N}\sum_{{\bf k}_1+{\bf k}_2+{\bf k}_3=\bf 0} 
\left(
J_{2{\bf k}_1}' c^\dagger_1c^\dagger_2a_{-3}
+J_{2{-\bf k}_1}' b^\dagger_1b^\dagger_2a_{-3}
+J_{2{\bf k}_3}' a^\dagger_1c_{-2}c_{-3}
+J_{2{-\bf k}_3}' a^\dagger_1b_{-2}b_{-3}
\right),\\
\label{h4'2}
{\cal H}_4' &=& \frac1N\sum_{{\bf k}_1+{\bf k}_2+{\bf k}_3+{\bf k}_4=\bf 0} 
\left(
J_{2{\bf k}_1+{\bf k}_3}' a^\dagger_1a^\dagger_2a_{-3}a_{-4}
+J_{2{\bf k}_2+{\bf k}_4}' a^\dagger_1c^\dagger_2a_{-3}c_{-4}
+J_{2{\bf k}_1+{\bf k}_3}' a^\dagger_1b^\dagger_2a_{-3}b_{-4}
\right.\nonumber\\
&&{}\left.
+ J_{2{\bf k}_2+{\bf k}_3}' b^\dagger_1c^\dagger_2b_{-3}c_{-4}
+ \frac12 J_{2{\bf k}_1+{\bf k}_4}' a^\dagger_1c^\dagger_2a_{-3}b_{-4}
+ \frac12 J_{2{\bf k}_2+{\bf k}_3}' a^\dagger_1b^\dagger_2a_{-3}c_{-4}
\right),
\end{eqnarray}
and $J_{2{\bf k}}' = 2J_2'\cos k_z(1+e^{-ik_x})$. We find from Eqs.~\eqref{h2} and \eqref{h2'2} for the spectrum of $\beta$-particles
\begin{equation}
\label{alpha2'}
\epsilon_\beta({\bf k}) = H+1-J+J\cos k_x - \cos\frac{k_x}{2}\left| 1-2J_2'\cos k_ze^{-ik_x} \right| - 2J_2'.
\end{equation}
One has from Eq.~\eqref{alpha2'} in the second order in $J_2'$ for the field value at which $\epsilon_\beta({\bf k})$ becomes gapless
\begin{equation}
\label{hc'2}
 H_c =	2J-1+\frac{1}{8J} + 2 J_2' + J_2'\frac{\left(1-8 J^2\right)}{16 J^3} + \left(J_2'\right)^2\frac{\left(1-4 J^2+8 J^4\right)}{16 J^5}.
\end{equation}
The spectrum $\epsilon_a({\bf k})$ of $a$-particles can be calculated in the second order in $J_2'$ as it is done for $J_1'$ in the main text. Considering diagrams shown in Fig.~\ref{disp} one obtains that $\epsilon_a({\bf k})$ has a minimum at ${\bf k}=(0,\pi)$ near which we have
\begin{eqnarray}
\label{ea'2}
\epsilon_a({\bf k}) &=& 2H+2-4J-\frac{1}{1+J}-4J_2' 
- \left(J_2'\right)^2 \frac{(1+4 J (1+J (1+J) (3+J)))}{2(1+J)^3 (1+2 J)^2} 
+ D_\| k_x^2 + D_\perp k_z^2,\\
\label{dperp2}
D_\perp &=& 
\left(J_2'\right)^2
\left(
\frac{J (1+J) (2 J (3+J (23+J (38+J (5+4 J ((J-1) J (5+J)-9)))))-5)-1}{16 J^5 (1+J)^4 (1+2 J)^2}
\right.\nonumber\\
&&{}
+\frac{(1+2 J) (1+4 J (J (\xi -1)-1)-\xi )}{64 \sqrt{2} J^5 (1+J)^4 \sqrt{\sqrt{\xi }-1+2 J^2}}
\nonumber\\
&&{}\left.
-\frac{4 \sqrt{2} \xi  (1-2 J (1+J) (1+2 J)+\xi )}{\sqrt{\xi \left(\sqrt{\xi }-1+2 J^2\right) } \left(40 J^2+J^3 (20-8 \xi )-2 (3+\xi )+J (3+\xi ) (4+\xi )\right)}
\right),
\end{eqnarray}
where $\xi =4 J^2 (5+4 J (2+J)) -3$, $D_\perp>0$ at $J>1.16$ and $D_\perp \approx 1/8J$ at $J\gg1$. One finds from Eq.~\eqref{ea'2}
\begin{equation}
\label{hs'2}
	H_s = 2J-1+\frac{1}{2+2J} + 2J_2' + \left(J_2'\right)^2 \frac{(1+4 J (1+J (1+J) (3+J)))}{4(1+J)^3 (1+2 J)^2} 
\end{equation}
that gives $H_s=47.08$~T in LiCuVO$_4$ in accordance with the value of 47.1~T obtained in Ref.~\cite{zhito}. We derive from Eqs.~\eqref{hc'2} and \eqref{hs'2} for the binding energy of two magnons given by $2(H_s-H_c)$ the value of $0.031J$ which is in good agreement with that of $0.030J$ obtained in Ref.~\cite{zhito}. Then, we find using Eqs.~\eqref{hc'2} and \eqref{hs'2} that the nematic phase can arise (i.e., $H_s>H_c$) if $|J_2'|<7.15$~K that is in accordance with the inequality $|J_2'|<6.97$~K obtained in Ref.~\cite{zhito}. It should be noted that small discrepancies in these values are attributed to the fact that the inter-chain interaction is taken into account exactly at $H\ge H_s$ in general equations which are solved in Ref.~\cite{zhito} numerically whereas our analytical results are obtained in the second order in $J_2'$.

It becomes very important at $H<H_s$ that the model we discuss is two-dimensional. General result for the condensate density in 2D Bose gas \cite{popov} reads in our notation at $H\approx H_s$ as
\begin{equation}
\label{rho'}
\rho = \frac{Z^2}{8\pi\sqrt{D_\|D_\perp}}(H_s-H)\ln\left( \frac{\sqrt{D_\|D_\perp}}{H_s-H} \right),
\end{equation}
where $Z$ is given by Eq.~\eqref{z}. Eq.~\eqref{rho'} is valid at 
\begin{equation}
\label{val2}
	H_s-H \ll \sqrt{D_\|D_\perp}.
\end{equation}
One obtains $\sqrt{D_\|D_\perp}\approx0.37$~T in LiCuVO$_4$ using Eqs.~\eqref{dpar} and \eqref{dperp2}. We find from Eqs.~\eqref{mag} and \eqref{rho'} for the static susceptibility
\begin{equation}
\label{chi}
	\chi(H) = \frac{d}{dH} \left(\frac1N\sum_j\left\langle S_j^z\right\rangle\right) = \frac{(1+J)^2 }{1+2 J} \frac{Z^2}{4\pi\sqrt{D_\|D_\perp}}\left(1+\ln\left( \frac{\sqrt{D_\|D_\perp}}{H_s-H} \right)\right).
\end{equation}
The static susceptibility has been measured recently \cite{svistov} in LiCuVO$_4$. One of the main results was an observation of a cusp in $\chi(H)$ upon entering into the nematic phase with the field increasing. Then, it was found that $\chi(H)\approx M_{sat}/2H_s$ in the nematic phase, where $M_{sat}$ is the saturation value of the magnetization per ion. This finding turns out to be in agreement with the prediction of Ref.~\cite{zhito} according to which $\chi(H)\approx 0.54M_{sat}/H_s$. 

Although Eq.~\eqref{chi} is valid quite close to $H_s$, when inequality \eqref{val2} holds, one can estimate $\chi(H)$ at $H_s-H\approx \sqrt{D_\|D_\perp}$ using Eq.~\eqref{chi}. We obtain from this equation by discarding the logarithm $\chi(H)\approx 0.40 M_{sat}/H_s$, where $M_{sat}=1/2$ is implied. This result agrees well with the experiment in view of the fact that Eqs.~\eqref{rho'} and \eqref{chi} are valid at $H_s-H\approx \sqrt{D_\|D_\perp}$ up to a constant of the order of unity.

It should be noted that a small interaction between chains making the system three-dimensional would screen the logarithm in Eqs.~\eqref{rho'} and \eqref{chi} and stabilize the long-range nematic order at finite $T$. Our finding that $H_s(0)-H_s(T)\propto T^{3/2}$, Eqs.~\eqref{eanem}, \eqref{teb}, \eqref{nemorder}--\eqref{mag} and \eqref{chizz} can be checked in further experiments on LiCuVO$_4$ which should confirm also that the phase observed just below $H_s$ is really the quadrupolar one.

\bibliography{nematic} 

\end{document}